\documentclass[preprint,12pt]{elsarticle}
\pdfoutput=1
\usepackage{multirow}
\usepackage{booktabs}
\usepackage{array}
\newcolumntype{L}[1]{>{\raggedright\arraybackslash}p{#1}}
\usepackage{float}
\usepackage{amssymb}
\usepackage[a4paper, top=2.5cm, bottom=2.5cm, left=1.9cm, right=1.7cm]{geometry}
\usepackage{amsmath}
\usepackage{cleveref}
\usepackage{soul} 
\usepackage[normalem]{ulem} 
\newcommand\redsout{\bgroup\markoverwith
{\textcolor{red}{\rule[0.5ex]{2pt}{0.4pt}}}\ULon}

\usepackage{xcolor}
\usepackage{threeparttable} 

\definecolor{myblue}{RGB}{0, 100, 200} 
\newcommand{\AM}[1]{\textcolor{black}{#1}}

\journal{Elsevier}

\begin{document}

\begin{frontmatter}

\title{Pull-off force prediction in viscoelastic adhesive Hertzian contact by physics augmented machine learning}

\author[inst1,inst2]{Ali Maghami}

\affiliation[inst1]{organization={Politecnico di Bari, Department of Mechanics Mathematics and Management},
            addressline={Via Orabona 4}, 
            city={Bari},
            postcode={70125}, 
            country={Italy}}

\author[inst2]{Merten Stender}
\author[inst1,inst3]{Antonio Papangelo \corref{cor1}}

\cortext[cor1]{Corresponding author:{ antonio.papangelo@poliba.it}}

\affiliation[inst2]{organization={Technische Universität Berlin, Chair of Cyber-Physical Systems in Mechanical Engineering},
            addressline={Straße des 17. Juni}, 
            city={Berlin},
            postcode={10623},
            country={Germany}}

\affiliation[inst3]{organization={Hamburg University of Technology, Department of Mechanical Engineering},
            addressline={Am Schwarzenberg-Campus 1}, 
            city={Hamburg},
            postcode={21073},
            country={Germany}}


\begin{abstract}
Understanding and predicting the adhesive properties of viscoelastic Hertzian contacts is crucial for diverse engineering applications, including robotics, biomechanics, and advanced material design. The maximum adherence force of a Hertzian indenter unloaded from a viscoelastic substrate has been studied with analytical and numerical models. Analytical models are valid within their assumptions, numerical methods offer precision but can be computationally expensive, necessitating alternative solutions. This study introduces a novel physics-augmented machine learning (PA-ML) framework as a hybrid approach, bridging the gap between analytical models and data-driven solutions, which is capable of rapidly predicting the pull-off force in an Hertzian profile unloaded from a broad band viscoelastic material, with varying Tabor parameter, preload and retraction rate. Compared to previous models, the PA-ML approach provides fast yet accurate predictions in a wide range of conditions, properly predicting the effective surface energy and the work-to-pull-off. The integration of the analytical model provides critical guidance to the PA-ML framework, supporting physically consistent predictions. We demonstrate that physics augmentation enhances predictive accuracy, reducing mean squared error (MSE) while increasing model efficiency and interpretability. We provide data-driven and PA-ML models for real-time predictions of the adherence force in soft materials like silicons and elastomers opening to the possibility to integrate PA-ML into materials and interface design. The models are openly available on Zenodo and GitHub.
\end{abstract}









\end{frontmatter}

\section{Introduction}
\label{sec:sample1}

Real-time insights into the adhesive properties of viscoelastic materials, such as polymers and elastomers, greatly benefit engineering applications in gripping technologies \cite{ji2019apple}, switchable adhesion \cite{liu2022switchable}, friction \cite{papangelo2024friction,ciavarella2020degree}, biomechanics \cite{guo2024interaction}, soft robotics \cite{giordano2024mechanochromic}, and climbing robotics technology \cite{tao2023climbing}. Predicting the adhesive properties in real time supports secure handling, adaptability to diverse surfaces, fine control \cite{liang2024autopeel}, and real-time material design \cite{humfeld2021machine}. Hertzian elastic adhesive contact lays its foundation in the theory derived by Johnson, Kendall and Roberts (JKR) \cite{johnson1971surface} which is valid for soft materials like silicons and elastomers, while for stiff contacts the approach of Derjaguin, Muller, and Toporov (DMT) is generally adopted \cite{derjaguin1975effect}. Tabor \cite{tabor1977surface} showed that the two approaches were both valid, but in the two different limits of short-range (JKR) and long-range (DMT) adhesion and that the transition from one to the other regime could be traced by varying a dimensionless parameter $\mu = \left( \frac{R \Delta \gamma_0^2}{{E_0^*}^ 2 h_0^3} \right)^{1/3}$, nowadays known as the "Tabor parameter", where $R$ is the radius of curvature of the Hertzian profile, $\Delta \gamma_0$ is the thermodynamic surface energy, $E_0^*$ is the plain strain elastic modulus and $h_0$ is the interatomic equilibrium distance. In 1992 Maugis \cite{maugis1992adhesion} derived the elastic adhesive contact solution for a Hertzian profile using a Dugdale potential, showing a smooth transition of the pull-off force $|P_{po}|$ (i.e. the maximum adherence force reached during unloading) from the DMT ($P_{po}=2\pi R\Delta\gamma_0$, valid for \AM{$\mu \lesssim 0.1$}) to the JKR regime ($P_{po}=(3/2)\pi R\Delta\gamma_0$, valid for \AM{$\mu\gtrsim3$}) \cite{greenwood1997adhesion, feng2000contact, papangelo2020numerical}. 

On the other hand, soft materials have viscoelastic behavior, and it has been shown that these have a huge effect in determining the pull-off force as rate-effects come at play \cite{tricarico2025enhancement,ciavarella2025dynamic, maghami2024viscoelastic, maghami2024bulk, afferrante2022effective,violano2021jkr}. Soft contact mechanics can be interpreted as a fracture mechanics problem \cite{johnson1971surface,maugis1992adhesion}: the energy required to propagate a viscoelastic crack must include not only the surface energy contribution but also the energy that is dissipated by the viscoelastic nature of the bulk material, which is rate-dependent \cite{schapery1975theory1, schapery1975theory2, greenwood1981mechanics, greenwood2004theory, schapery2022theory, persson2005crack, persson2017crack, persson2021simple, carbone2022theory, mandriota2024adhesive}. The macroscopic result is that the pull-off force can be potentially amplified by a factor \AM{$E_\infty/E_0={1}/{k}$} \cite{greenwood2004theory, persson2005crack, papangelo2023detachment, maghami2024viscoelastic, maghami2024bulk}, which is commonly of the order of $10^3$ in soft silicons \cite{maghami2024bulk} ($E_\infty$ and $E_0$ being respectively the glassy and the rubbery Young modulus of the material and $k$ is called the modulus ratio). The theory of Persson and Brener \cite{persson2005crack} demonstrated to be accurate in predicting the rate-dependent pull-off force in \textit{linear viscoelastic solids} and it can easily incorporate the case of real materials with broad-band viscoelastic spectrum \cite{maghami2024bulk}. However, it has some limitations: it does not account for boundary effects \cite{violano2022size, maghami2024viscoelastic, maghami2024bulk}, it is valid only in the short-range adhesion limit (JKR) \cite{wang2025rapid, violano2022long}, it assumes that the surrounding material is fully relaxed and dissipation is limited in a zone close to the contact edge \cite{afferrante2022effective,violano2021jkr}\AM{, hence it requires a minimum preload to be accurate \cite{violano2022size}.}

To overcome these difficulties, numerical modeling of viscoelastic adhesive systems offers valuable tools for computing the adhesive properties, requiring the definition of a relaxation function \cite{maghami2024bulk} and the use of time-marching methods, which involve the use of iterative schemes \cite{ahmad2024family} to solve the nonlinear contact problem \cite{papangelo2020numerical, papangelo2023detachment}. As a result, material characteristics, compliance, and loading properties heavily influence the efficiency of numerical simulation \cite{souza2010multiscale}. This poses limitations in the use of numerical models for real-time predictions of the contact state and limits the possibility to integrate contact analysis into the optimization loop that leads to machine design. Real-time predictions through numerical methods face challenges related to computational scaling, as compute time varies widely across scenarios, and strictly depends on the physics to resolve, irrespective of the numerical approach taken \cite{souza2010multiscale}. 

In recent years, machine learning (ML) has emerged as a powerful tool for addressing complex challenges, offering data-driven methods for designing materials with speed and acceptable precision \cite{guo2021artificial}. ML offers a powerful framework \cite{Goodfellow-et-al-2016} for uncovering patterns and relationships in high-dimensional, large-scale datasets, particularly in cases where analytical models are inadequate \cite{mackay2023informed} or when evaluating numerical models has a high computational cost \cite{mackay2023informed}. While ML-based models represent approximators of finite accuracy, they can provide a more computationally efficient alternative to physics-based approaches. Notably, although the computational cost of ML models generally scales with model complexity and the dimensionality of inputs and outputs, the inference time remains constant once the model is trained \cite{mackay2023informed}. As a result, the inference latency is unaffected by the variability in input data, potentially encoding different complexity in the underlying approximation task, resulting in a fixed computational time.

ML and neural network-based approaches have found remarkable applications in the fields related to materials \cite{zhu2024transfer, javadi2022deep, kellner2019establishing, eshkofti2024modified}, fracture \cite{wang2021machine, athanasiou2023integrated, yi2024mechanics, perera2023generalized, li2022machine}, contact mechanics \cite{goodbrake2024neural, motiwale2024neural, kalliorinne2021artificial, sahin2024solving, sahin2024physics}, and tribology \cite{stender2021deep, geier2023machine, sattari2020prediction} and mechanics \cite{didonna2019reconstruction}. More related to adhesion, ML research has trended toward designing adhesive pillar geometry, initiated by Kim et al. \cite{kim2020designing}, who explored elastic behavior in pillar designs, and tuning the stress at the flat contact region. This idea has been extended to studies on PDMS and elastomers, supported by experimental validation \cite{son2021machine, luo2022machine, kim2023designing, dayan2024machine}. More recently a ML-based design approach focused on fibrillar adhesives has been presented \cite{shojaeifard2025machine}. Despite significant advancements, the works mentioned above primarily focus on flat-to-flat contacts, utilizing data and ML as problem-specific surrogate models to optimize non-contacting parts. Furthermore, while ML offers notable advantages, it often comes at the cost of losing physical concepts within the model and relying solely on data. We aim to bridge these gaps by leveraging physics-based analytical solutions to guide ML models in gaining insights into the Hertzian viscoelastic adhesive contact problem.

\begin{figure}[H]
\centering
\includegraphics[width=0.75\textwidth]{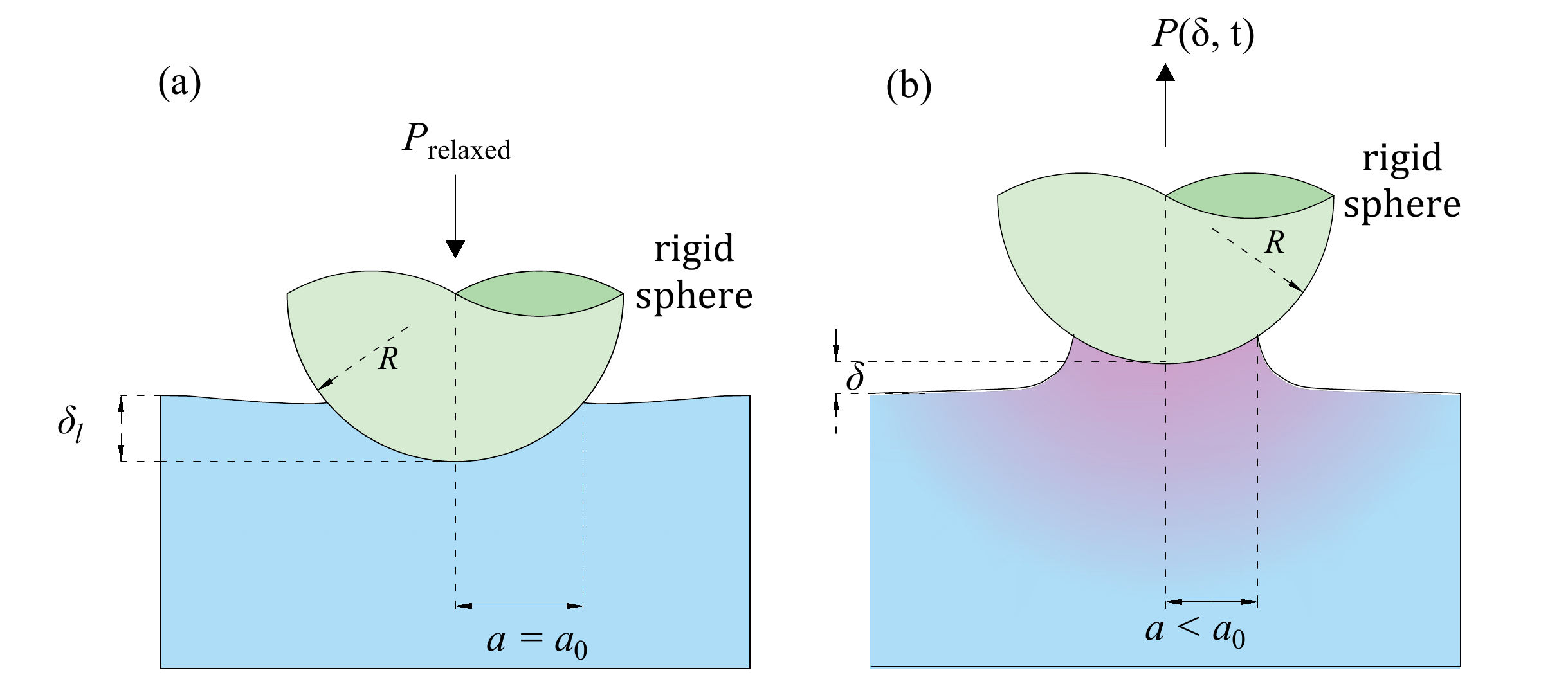}
\caption{Schematic representation of the contact interaction between a rigid sphere and a viscoelastic surface: \AM {(a) viscoelastic surface in the fully relaxed state under contact, (b) unloading phase at a constant unloading rate.}}\label{fig:indenter}
\end{figure}

In this study, we address the problem of a rigid sphere with radius \( R \) being unloaded from a relaxed broad-band viscoelastic adhesive half-space (see Fig.~\ref{fig:indenter}). We aim to present ML-based models, both classical (ML) and physics-augmented (PA-ML), that predict the pull-off force and the work to pull-off as a function of \AM{five} parameters: the Tabor parameter $\mu$, the exponent $n$ characterizing the broadness of the material spectrum, \AM{the modulus ratio $k$,} the normalized indentation depth reached during the quasi-static loading phase, \AM{and} the normalized unloading rate. A detailed comparative analysis will be conducted to evaluate the accuracy of different ML algorithms (Linear Regression, Regression Tree, Random Forest, and XGBoost will be considered) and to compare the performance of a pure data-driven approach (ML) against that of a physics-augmented ML model (PA-ML).

The remainder of the paper is structured as follows: Section \ref{sec:BEM} provides a detailed description of our numerical model, including the normalization approach that enables comparison with analytical models. This section also includes a discussion on the scaling of input parameters and their impact on the computational cost of the numerical model; Section \ref{sec:ML} presents \AM{the range of exploration as well as} the background on ML and describes the ML models and PA-ML models employed in this study; \AM{Section \ref{sec:results} provides the results of ML and PA-Ml models;} Finally, Section~\ref{sec:conc} summarizes the main findings and presents concluding remarks.

\section{Physical background and simulation framework} \label{sec:BEM}
\begin{figure}[t]
\centering
\includegraphics[width=0.9\textwidth]{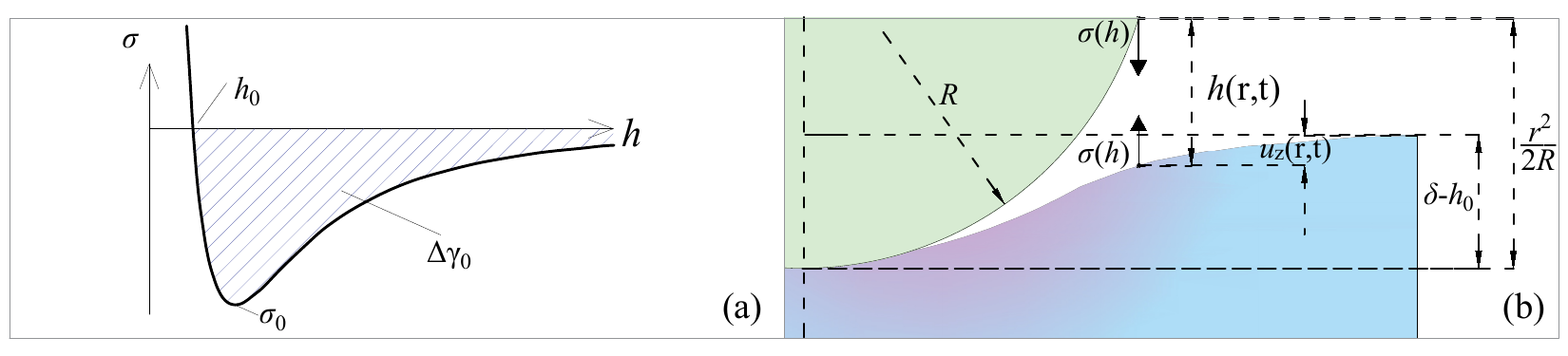}
\caption{(a) The Lennard-Jones law depicting the adhesive stress $\sigma(h)$ as a function of the gap $h$. 
(b) The geometric components of the gap function $h(r,t)$, are shown for an arbitrary surface point, including contributions of indentation ($\delta$), equilibrium distance ($h_0$), the parabolic approximation of the indenter geometry ($r^2/R$ where $R$ is the radius of the sphere), and viscoelastic deformation at an arbitrary point ($u_z(r,t)$).}\label{contact_surface}
\label{fig:contact_surface}
\end{figure}

 Let us consider a rigid sphere of radius $R$ that is loaded against a viscoelastic adhesive halfspace up to an initial indentation depth $\delta_l$ and later is unloaded at a constant unloading rate $r_u$ (see Fig. \ref{fig:indenter}). In the following we will assume that unloading starts from a given indentation and that the substrate material is in a fully relaxed state (Fig. \ref{fig:indenter}(a)). The interaction at the interface of the rigid sphere and the viscoelastic halfspace is characterized by the Lennard-Jones force-separation law \cite{johnson1997adhesion}, as:
\begin{equation}
\sigma\left(  h\right)  =-\frac{8\Delta\gamma_{0}}{3h_{0}}\left[  \left(
\frac{h_{0}}{h}\right)  ^{3}-\left(  \frac{h_{0}}{h}\right)  ^{9}\right]\;,
\label{eq:LJ}%
\end{equation}
where \(\sigma\) is defined as the interfacial stress (\(\sigma > 0\) for compression), \(h\) is the local gap, \(h_{0}\) is the equilibrium distance (also called the interatomic spacing, with $\sigma(h_0)=0$), and \(\Delta\gamma_{0}\) is the thermodynamic surface energy given by \(\Delta\gamma_{0} = \frac{9\sqrt{3}}{16}\sigma_{0}h_{0}\), where $\sigma_0$ as depicted in Figure \ref{fig:contact_surface}(a), is the maximum tensile stress that happens at $h=3^{(1/6)}h_0$ \cite{greenwood1997adhesion}.
The sphere is approximated by a Hertzian profile with radius of curvature $R$. The gap function in the direction normal to the nominal contact plane, at the radial coordinate $r$ at time $t$ (Fig. \ref{fig:contact_surface}b) is expressed as:
\begin{equation}
h(r,t)=-\delta(t)+h_{0}+\frac{r^{2}}{2R}+u_{z}\left(  r,t\right)\;,  \label{eq:h}%
\end{equation}
where \(\delta\) is the indentation which is positive as the sphere approaches the viscoelastic half-space, and \(u_{z}(r,t)\) represents the deflection of the viscoelastic half-space, which depends on the loading history (showing the dependence of the gap function on time \(t\)). According to the elastic-viscoelastic correspondence principle expressed through Boltzmann integrals \citep{christensen2012theory}, and considering the vertical deflections of the halfspace for an elastic axisymmetric problem \cite{greenwood1997adhesion, feng2000contact}, the normal displacements \(u_{z}(r,t)\) of the viscoelastic half-space at time \(t\) and position \(r\) will be dependent on the history of contact, as follows:
\begin{equation}
{u_{z}(r,t)=\int G\left(  r,s\right)  s\int_{-\infty
}^{t}C(t-\tau)\frac{d\sigma(s,\tau)}{d\tau}d\tau ds}\;, \label{eq:uz}%
\end{equation}
where $G\left(  r,s\right)$ is the Kernel function which is given in \ref{app:BEM}, and $C(t)$ is the creep compliance function \cite{maghami2024bulk}. \AM{To capture a wide range of broad-band viscoelastic behavior, we adopt the modified power-law model for viscoelastic materials} \cite{williams1964structural}, and the creep compliance function is given as \cite{maghami2024bulk}:
\begin{equation}
C(t) = C_{0} - 2 \frac{(C_{0} - C_{\infty})}{\Gamma(n)} \left(\frac{t}{\tau_{0}}\right)^{n/2} \mathbf{K}_{n} \left(2 \sqrt{\frac{t}{\tau_{0}}}\right) \,, \label{eq:Ctgen}
\end{equation}
where \(\mathbf{K}_{n}(x)\) is the modified Bessel function of the second kind, \AM{$\tau_0$ is the relaxation time of the material,} $\Gamma(n)$ represents the Gamma function, $C_0=1/E_0$, $C_\infty=1/E_\infty$, with $E_0$ and $E_\infty$ respectively the rubbery and glassy moduli.\AM{For \( n \approx 1.6 \), the model reproduces behavior similar to the Standard Linear Solid (SLS), a classical viscoelastic model composed of a spring in parallel with a Maxwell element (a spring and dashpot in series).
}

Incorporating Equations (\ref{eq:LJ}) and (\ref{eq:Ctgen}) into (\ref{eq:uz}), and combining them with (\ref{eq:h}) gives a nonlinear convolutional integral equation for the unknown $h(r,t)$. Details regarding the numerical implementation of the Boundary Element Method (BEM) numerical model and the normalized form of the Equations (\ref{eq:LJ}), (\ref{eq:h}), (\ref{eq:uz}), and (\ref{eq:Ctgen}) can be found in \ref{app:BEM}.

\begin{figure}
\centering
\includegraphics[width=0.47\textwidth]{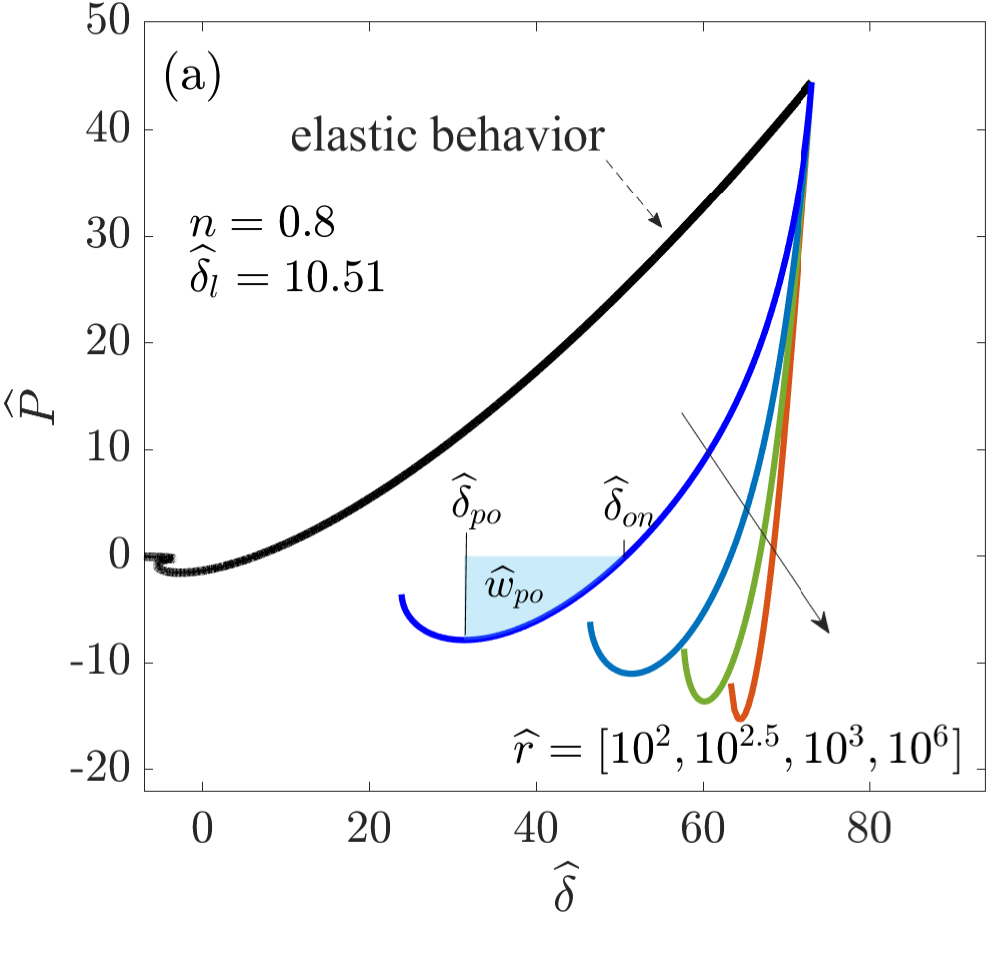}
\includegraphics[width=0.47\textwidth]{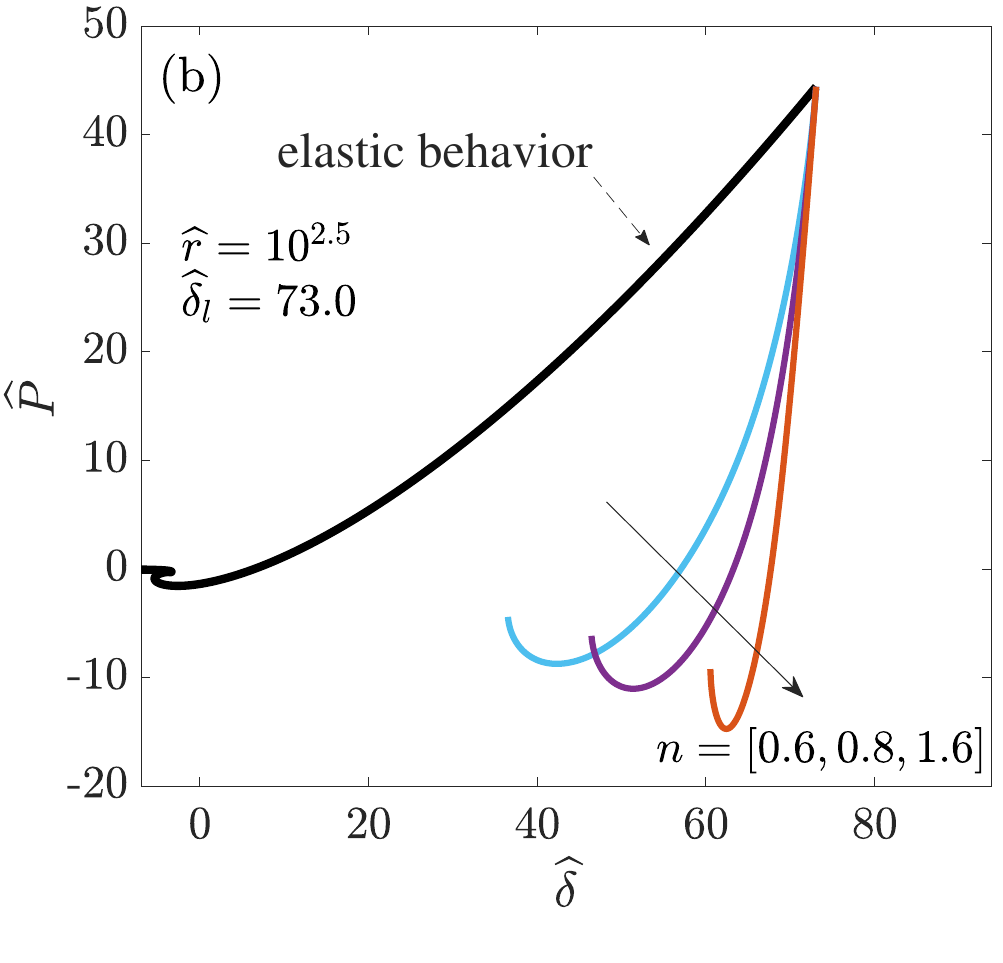}
\includegraphics[width=0.47\textwidth]{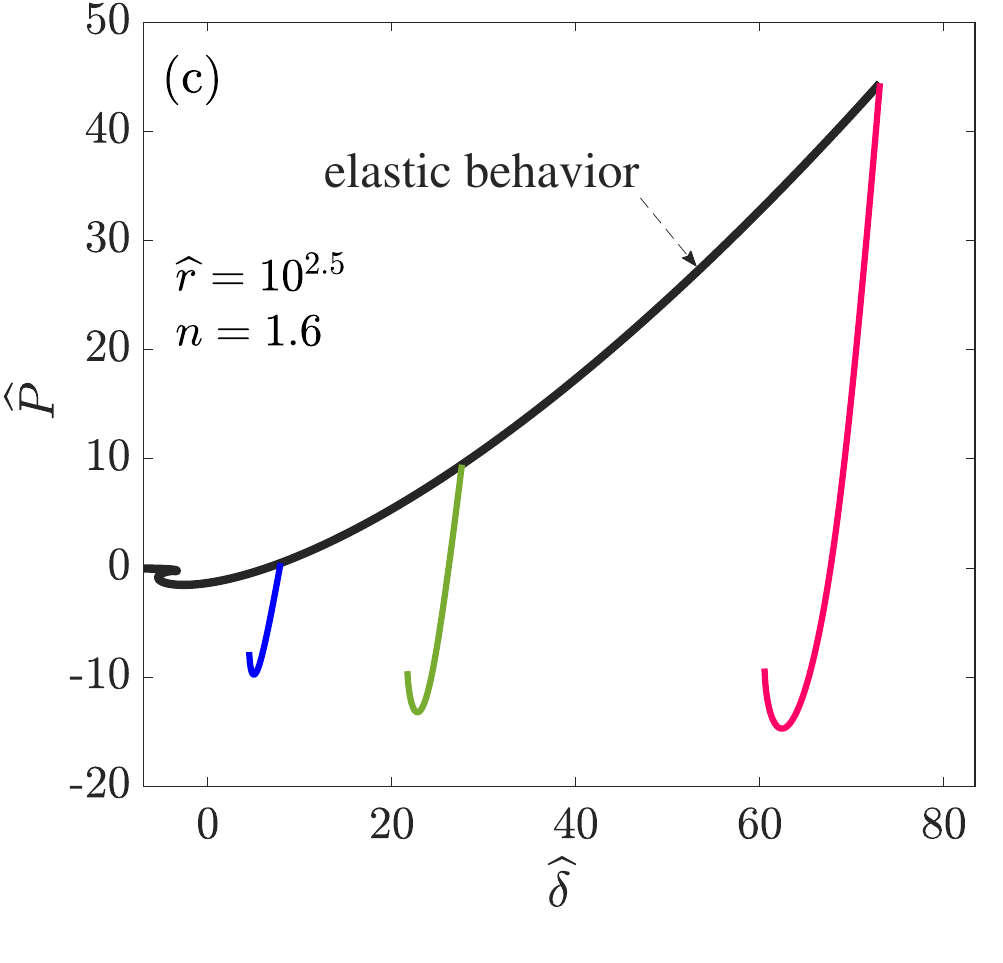}
\includegraphics[width=0.47\textwidth]{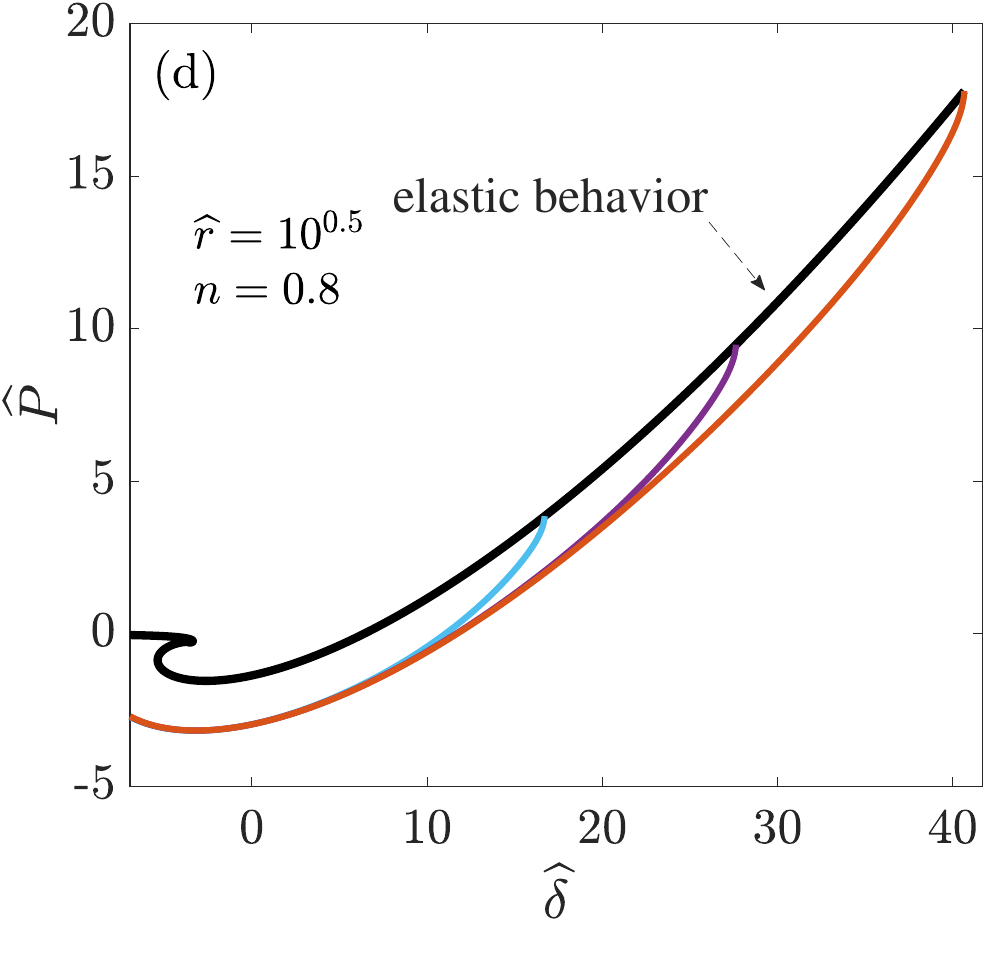}
\caption{Load vs. indentation curves demonstrating adhesive interactions (with negative load values indicating tensile forces) between a Hertzian indenter and a viscoelastic substrate. All panels correspond to a fixed Tabor parameter of $\mu = 3.24$ and dimensionless modulus ratio of $k = 0.1$.
(a) Effect of unloading rate on the adhesive response, for a fixed initial indentation depth of $\widehat{\delta_l}=73.0$ and material power-law exponent $n=0.8$;
(b) Influence of the material’s power-law exponent $n=[0.6,\,0.8,\,1.6]$, under fixed initial indentation depth of $\widehat{\delta_l}=73.0$ and unloading rate of $\widehat{r}=10^{2.5}$;
(c) Effect of different initial indentation depths, $\widehat{\delta}_l = [7.6, 27.8, 73.0]$, with constant material exponent of $n=1.6$ and unloading rate of $\widehat{r}=10^{2.5}$;
(d) Observation of saturation behavior for $\widehat{r}_u = 10^{0.5}$, $n = 0.8$, and $\widehat{\delta}_l = [16.7,\,27.6,\,40.7]$, where the pull-off force approaches a limiting value beyond a critical indentation depth, indicating the presence of a threshold in the adhesive response.}

\label{fig:load_vs_disp}
\end{figure}

In the normalized formulation of the problem, the inputs to the numerical model can be reduced to: the Tabor parameter $\mu$, the dimensionless parameters $ \{k=E_0/E_\infty,n \}$
that define the viscoelastic properties of the material, the normalized indentation depth at the beginning of the unloading phase $\widehat{\delta_l}={\delta_l}/{h_0}$, and the normalized unloading rate $\widehat{r_u}={r_u}/{\tau_0 h_0}$. The output of the numerical model includes the interfacial gap $h$, the pressure distribution, the normal force $P$, the indentation $\delta$ as a function of time. Figure \ref{fig:load_vs_disp} shows the dimensionless normal force $\widehat{P}=\frac{P}{1.5\pi\Delta\gamma_0 R}$ as a function of the dimensionless indentation $\widehat{\delta}$ for different combinations of the input parameters $\{\widehat{\delta_l},n,\widehat{r_u}\}$ for $\mu=3.24$ and $k=0.1$. These curves illustrate the relationship between the applied load and the indentation depth under various input conditions. One can observe the influence of the normalized input parameters, such as the unloading rate (Figure \ref{fig:load_vs_disp}(a)), the power law exponent (Figure \ref{fig:load_vs_disp}(b)), and the indentation-depth (Figure \ref{fig:load_vs_disp}(c) and (d)), on the force-displacement behavior and its effect on key parameters such as the maximum adherence force reached during unloading $\widehat{P_{po}} = |\min(\widehat{P})|$ (known as pull-off force) and work to pull-off which is defined as \AM{$
w_{\text{po}} =  \int_{{\delta}_{\text{on}}}^{{\delta}_{\text{po}}} {P}(\delta, t) \, d{\delta} ={\widehat{w}_{po}}{(1.5 \pi \Delta\gamma_0Rh_0)},$}
where ${\delta}_{\text{on}}$ denotes the displacement at which the normal force first becomes zero during unloading (i.e., the onset of tensile loading), and ${\delta}_{\text{po}}$ is the displacement at pull-off \cite{violano2022size}. This quantity represents the area under the unloading curve in the tensile regime and captures the energy dissipated during detachment, including both adhesive and viscoelastic contributions \AM{as it is highlighted in Figure \ref{fig:load_vs_disp}(a).} 

While the BEM model provides detailed insight into the adhesive response under a wide range of conditions, its nonlinear and history-dependent nature results in significant computational demands. Specifically, the total computational cost scales with the product of spatial discretization size, number of time steps, and convergence iterations per step—making real-time prediction across large parameter spaces challenging. To complement these detailed simulations, we incorporate an analytical model recently developed by Maghami et al.~\cite{maghami2024bulk}, which extends the classical steady-state theory of Persson and Brener (PB) to accommodate \AM{broad-band viscoelastic behavior through} modified power-law (MPL) viscoelastic materials. Referred to here as the extended PB (XPB) model, this closed-form formulation offers an efficient means of estimating the effective surface energy during adhesive crack propagation. In normalized form, the effective surface energy is approximated as \AM{$\widehat{\Gamma}_{\text{eff}} = \frac{\Delta \gamma_{\text{eff}}}{\Delta \gamma_0} \approx \frac{P_{\textrm{po}}}{1.5\pi\Delta\gamma_0R}={\widehat{P}_{\text{po}}}$},
where \( \widehat{P}_{\text{po}} \) is the \AM{normalized} pull-off force. The XPB model expresses \( \widehat{\Gamma}_{\text{eff}} \) as a function of crack velocity $v$ and viscoelastic material parameters through the integral:
\begin{equation}
\widehat{\Gamma}_{\text{eff}} = \left[ 1 - \left( 1 - k \right) \int_{0}^{+\infty} \frac{\widehat{\tau}^{n-1} \exp \left( -\widehat{\tau} \right)}{\Gamma(n)} \left[ \sqrt{1 + \left( \frac{\widehat{\Gamma}_{\text{eff}}}{2 \pi \widehat{v}} \frac{1}{\widehat{\tau}} \right)^{2}} - \left( \frac{\widehat{\Gamma}_{\text{eff}}}{2 \pi \widehat{v}} \frac{1}{\widehat{\tau}} \right) \right] d\widehat{\tau} \right]^{-1}\;,
\label{eq:PBmodel}
\end{equation}
with the normalized parameters defined as $
\widehat{v} = \frac{v \tau_{0}}{l_{0}}$, $\widehat{\tau} = \frac{\tau}{\tau_{0}}$,
where \( l_0 = \frac{E_{0}^{*} \Delta \gamma_{0}}{\pi \sigma_{c}^{2}} \) is a stress-based characteristic length and the critical stress \AM{\( \sigma_c=\alpha\sigma_0 \)} is related to the peak tensile stress \( \sigma_0 \) from the Lennard-Jones interaction through a proportionality factor \( \alpha \approx \pi / 9 \), as shown in~\cite{maghami2024bulk}. Furthermore, the crack velocity is shown to have the following relation with the unloading rate, $\widehat{v}=2.887\widehat{r}_u^{1.171}$,
thus, the XPB can be viewed as a 3-input model that takes $k$, $n$, and $\widehat{r}_u$ as inputs, then gives $\widehat{\Gamma}_{eff}$ as the output.

\begin{figure}[H]
\centering
\includegraphics[width=0.9\textwidth]{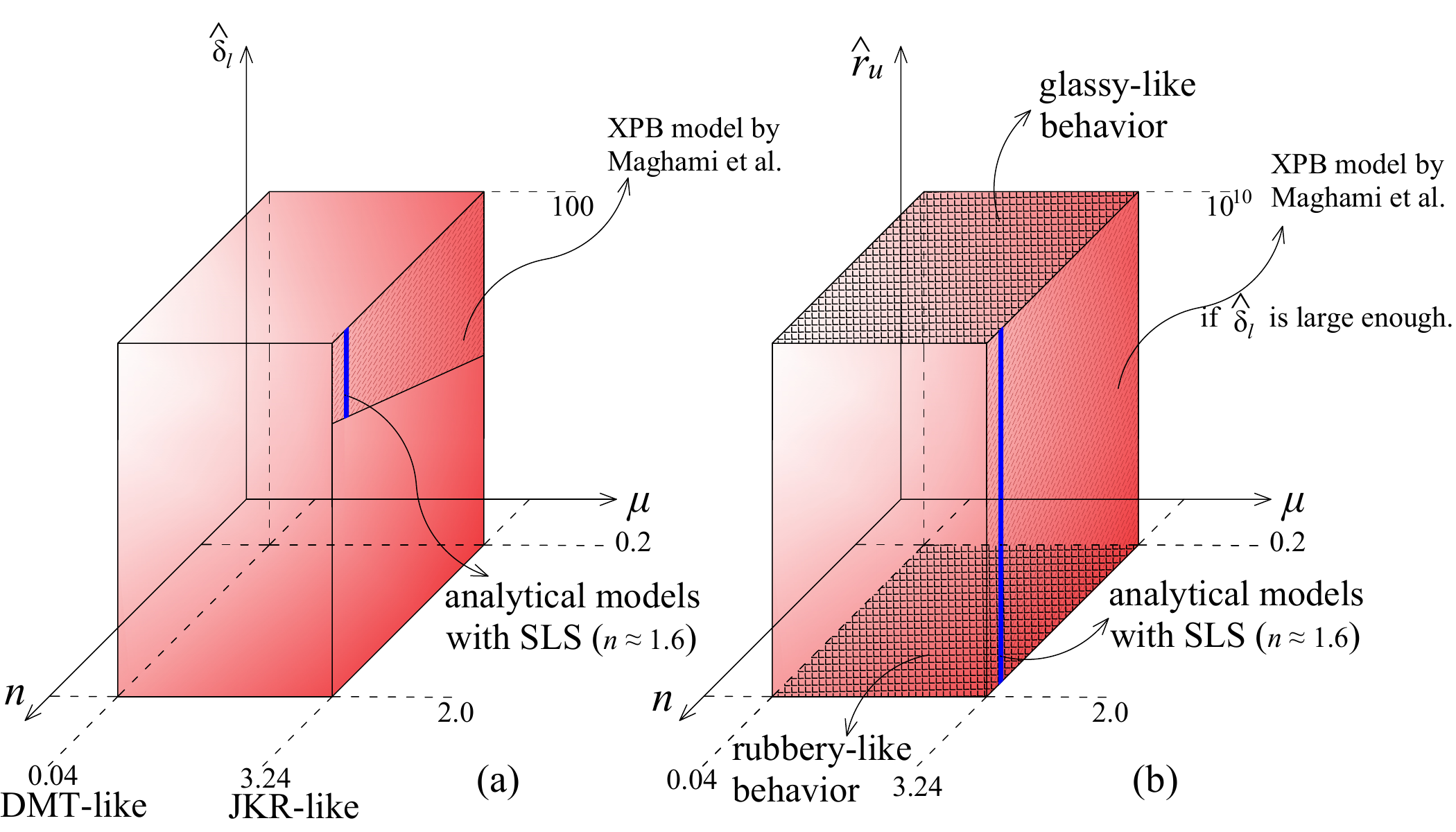}
\caption{Visualization of the material exploration space represented schematically through cuboid shapes in a multi-dimensional parameter space. The figure highlights specific regions: the glassy-like and rubbery-like regions (checkered pattern), the XPB analytical model region (hatched plane), and the analytical models based on Standard Linear Solid (SLS shown as blue line). The space is color-coded to qualitatively indicate computational cost, with intense color representing high computational expense and lighter shades indicating easily computable configurations: (a) Hypercuboid in the $(n,\, \mu,\, \widehat{\delta}_l)$ space and (b) hypercuboid in the $(n,\, \mu,\, \widehat{r}_u)$ space.} \label{fig:design_Space}
\end{figure}
\section{Machine learning for generalization beyond analytical models} \label{sec:ML}
\subsection{Range of exploration beyond analytical models}
Figure~\ref{fig:design_Space} schematically visualizes the parameter space explored in this work as three-dimensional cubes. The left cuboid spans the power-law exponent $n$, the Tabor parameter $\mu$, and the normalized initial indentation depth $\widehat{\delta}_l$. In this study, the exponent $n$ ranges from wide band behavior of $n = 0.2$ encompassing the range commonly observed in silicone-based polymers at room temperature (e.g., PDMS, as reported in \cite{shintake2018soft, sahli2019shear}) to values around $2.0$ related to a very narrow banded behavior. The Tabor parameter $\mu$ ranges from $0.04$ related to DMT-like to $3.24$ JKR-like behavior, while $\widehat{\delta}_l$ extends up to 100. The color gradient in this \AM{cuboid} qualitatively encodes computational cost, with deeper red tones indicating higher expense, particularly for large $\widehat{\delta}_l$ and \AM{large} $\mu$ \AM{which implies fine spatial and tempral discritization as well as extended time span.}. The blue vertical line at $n \approx 1.6$ identifies the region where SLS analytical models remain valid. The hatched surface indicates the domain in which the XPB model is applicable, specifically in the saturated indentation regime. The right cuboid in Figure \ref{fig:design_Space} spans $n$, $\mu$, and the normalized unloading rate, which  $\widehat{r}_u$ varies from very slow unloading of ($\widehat{r}_u = 10^{-1.5}$) close to pure rubbery like behavior to extremely rapid unloading ($\widehat{r}_u = 10^{10}$) to capture the glassy behavior. The hatched plane face marks the validity of the XPB model \cite{maghami2024bulk} when $\widehat{\delta}_l$ is sufficiently large (not shown in this projection), and the blue vertical line again indicates where the SLS approximation holds across all values of $\mu$ and $\widehat{r}_u$. However, substantial portions of the parameter space remain outside the validity limits of \AM{XPB} \cite{maghami2024bulk}.
\subsection{Machine learning framework for tabular data}
\begin{figure}[h]
\centering
\includegraphics[width=1.0\textwidth]{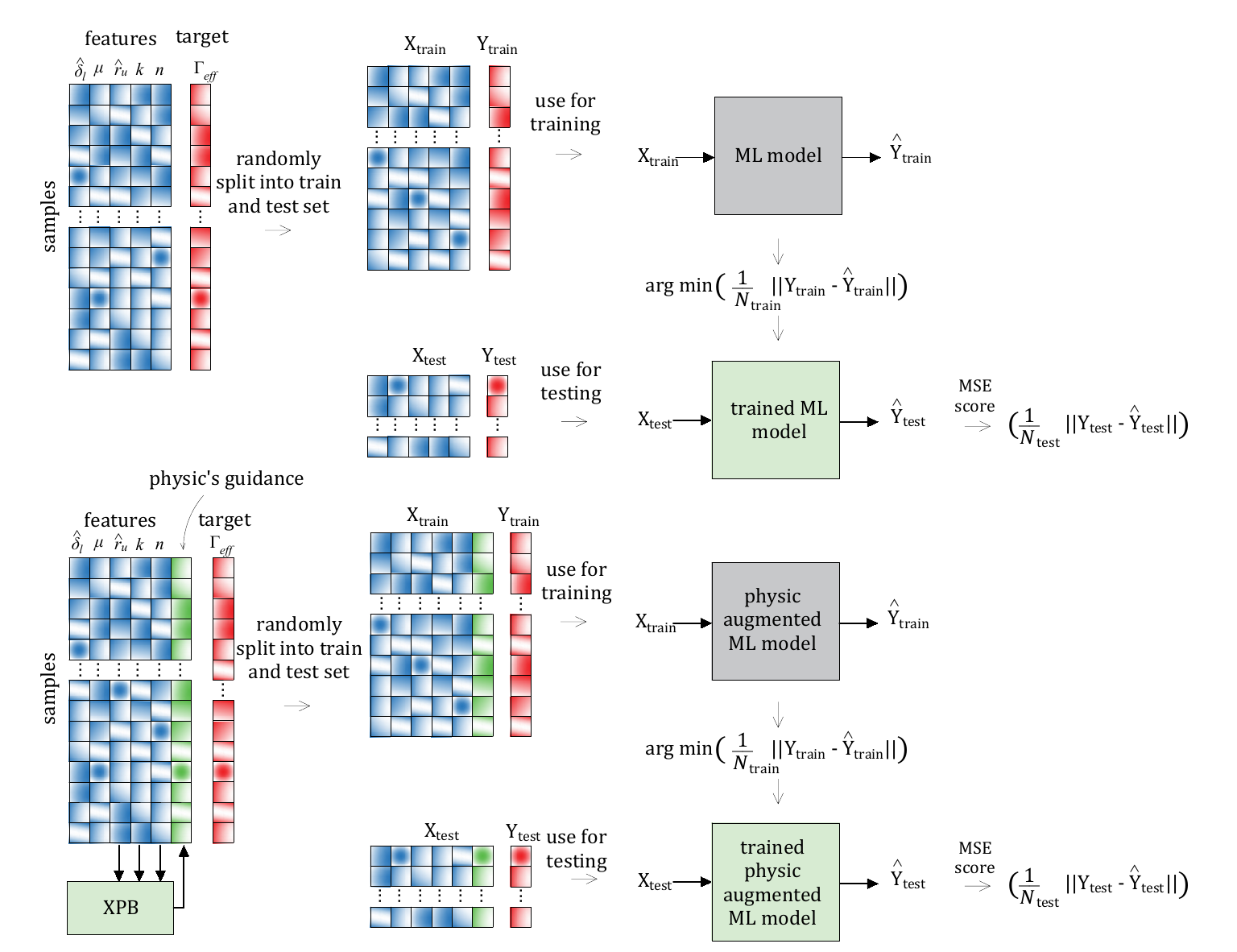}
\caption{Schematic representation of the data processing and modeling workflow. The tabular data are partitioned into train and test datasets. The standard model is trained using the training set by minimizing the MSE objective function. The physics-augmented model incorporates the analytical XPB framework during training to enhance model accuracy and generalization. The performance of each model is evaluated on the test set using MSE as the scoring metric.} \label{fig:fig_tab_data}
\end{figure}

ML for tabular data applies algorithms to structured datasets organized in rows (samples) and columns (features) to analyze and predict outcomes. In supervised learning, models trained on labeled data map inputs to outputs, predicting continuous values for regression tasks by minimizing errors, often using Mean Squared Error (MSE) loss. Data is split into training, validation, and test sets: the training set is used for model parameter optimization, validation set ensures overfitting preventions. Finally, the test set evaluates the model's performance on unseen data, providing an unbiased assessment of its predictive capabilities. This structured approach promotes effective learning and reliable performance in practical applications. The data processing and modeling workflow employed in this study is outlined in Figure \ref{fig:fig_tab_data}. 

The work at hand is characterized by tabular data, as described in Section~\ref{sec:BEM} where one can see the inputs as the Tabor parameter $\mu$, material exponent $n$, modulus ratio $k$, normalized indentation depth $\widehat{\delta_l}$, normalized unloading rate $\widehat{r}_u$ and output parameters would be the key characteristics of pull off ($\widehat{\Gamma}_{eff}$ or $\widehat{w}_{po}$). Moreover, our work faces the challenge of small and sparse data due to the computational efforts related to obtaining numerical solutions in regimes where the XPB model is valid. The \AM{total} number of data samples used in this work does not exceed {8505}, making the use of classical ML models efficient and practical. The descriptive statistics of the dataset, including measures of central tendency, variability, and data transformations, are provided in \ref{app:descriptive_stats}.

In this section, for the prediction of effective surface energy and the work to pull off, we utilize a linear regression model as a trivial baseline and tree-based ensemble methods for building the data-driven predictive models. Specifically, we compare regression trees, random forest regressors, and Extreme Gradient Boosting (XGBoost) models using k-fold cross-validation. See \ref{app:ML} for more details on each of the ML models. We note that physical augmentation can be achieved in various ways \cite{mackay2023informed}. In this work, we use the analytical XPB model to make predictions based on the parameters $k$, $n$, and $\widehat{r}_u$. These analytical results are fed as an additional input to the data-driven model. Thereby, the data-driven model has to learn to compensate for the modeling error of the analytical formulation in regimes where the analytical model is inadequate. The training process of the physics-augmented model, incorporating data augmentation through XPB, is depicted in Figure \ref{fig:fig_tab_data}.
\section{Results}\label{sec:results}
\subsection{Results for effective surface energy prediction} \label{sec:RES}
This section compares two model families for predicting viscoelastic behavior. The first is a data-driven model with five input parameters (denoted as ML), while the second incorporates an additional input from the analytical XPB model, making it a physics-augmented (denoted by PA-ML) model with six inputs. Both models are trained using Linear Regression, Regression Tree, Random Forest, and XGBoost. Their performance is assessed via 5-fold cross-validation, and the best model is tested on unseen data. The subsections cover the data-driven model first, followed by the physics-augmented model.

Table~\ref{tab:ML_results} summarizes the performance of various ML models in terms of MSE and $\textrm{R}^2$ values (mean ± standard deviation), with MSE scaled by $10^{-3}$, along with model sizes, i.e. number of trainable parameters. The results in Table~\ref{tab:ML_results} highlight the trade-offs between performance and model complexity across the four ML models. Linear Regression exhibits the highest MSE and the lowest \(\textrm{R}^2\) under cross-validation, indicating limited predictive accuracy. In contrast, ensemble-based models, such as Random Forest and XGBoost, achieve significantly lower MSEs and higher \(\textrm{R}^2\) values, with XGBoost showing the best performance. We note that this superior performance comes with increased model complexity. However, XGBoost has a smaller model size compared to Random Forest, despite its excellent predictive capability. \AM{More details on the performance of the data-driven models are given in \ref{app:ML}.}

\begin{table}[h!]
    \caption{Performance of ML approaches for the prediction of effective energy surfaces: MSE and $\textrm{R}^2$ values (mean ± std. deviation), with MSE reported as $\cdot 10^{-3}$ obtained from 5-fold cross-validation model and model sizes as number of trainable parameters included.} \label{tab:ML_results}
    \centering
    \small
    \begin{tabular}{lccc}
        \toprule
         & \textbf{MSE ($\cdot 10^{-3}$)} & \textbf{$\textrm{R}^2$} & \textbf{model size} \\
        \midrule
Linear Regression & $19.1948 \pm 0.7166$ & $0.8384 \pm 0.0047$ & 6\\
Regression Tree & $0.3502 \pm 0.0996$ & $0.9970 \pm 0.0008$ & 6457\\
Random Forest & $0.2155 \pm 0.0874$ & $0.9982 \pm 0.0007$ & 416700\\
XGBoost & $0.1449 \pm 0.0411$ & $0.9988 \pm 0.0003$ & 4993\\
        \bottomrule
    \end{tabular}
 
\end{table}

\begin{figure}[h] 
     \centering
      \includegraphics[width=0.49\textwidth]{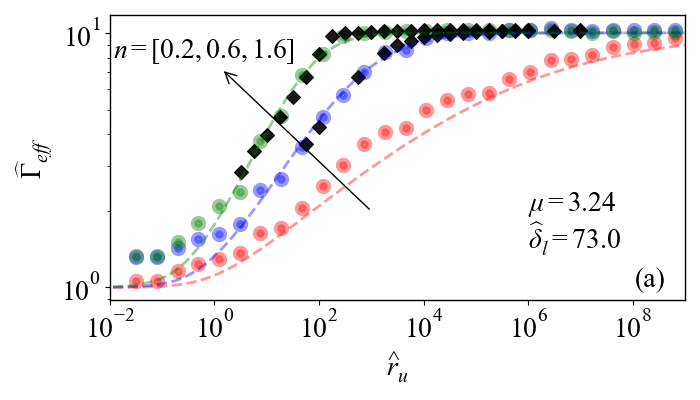}
      \includegraphics[width=0.49\textwidth]{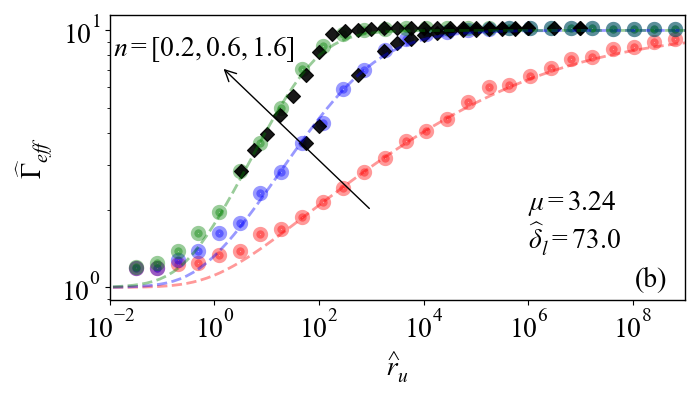}
              \caption{Comparison of the predictions from the purely data-driven machine learning (ML) approach (XGBoost) and the physics-augmented machine learning (PA-ML) approach (Random Forest) for the relationship between normalized surface energy and unloading rate. The dashed lines represent the XPB solution, diamond markers indicate the training data, and circles denote the predictions from the ML models. The models are evaluated for the parameters $\mu = 3.24$, $\widehat{\delta}_{\text{load}} = 73$, $k = 0.1$\AM{, and for three values of $n=[0.2, 0.6, 1.6]$}: (a) results from the purely data-driven ML model (\AM{XGBoost had the best performance}) and (b) results from the PA-ML model (\AM{Random Forest had the best performance}).} \label{fig:ML_surface}
\end{figure}

The \AM{best-performed model} (XGBoost) is utilized to predict the effective surface energy within the training data regime, but also outside that regime. This approach aims to evaluate the model's out-of-sample generalization. As discussed in Section~\ref{sec:BEM}, the computational costs of obtaining numerical samples for high values of indentation depth and {large} Tabor parameter are substantial. Additionally, the numerical model is sensitive to low values of \(n\). For instance, when \(n \leq 0.2\), our computational model fails to determine the pull-off force for cases where \(\widehat{\delta}_l \geq 73\) and \(\mu \geq 3.24\) at a reasonable computational cost. Figure~\ref{fig:ML_surface}(a) displays the \AM{ML}-predicted results for effective surface energy (represented by circle dots) compared with the XPB model solution (depicted with dashed lines). Notably, Figure~\ref{fig:ML_surface} represents a specific subset of the parameter space, specifically for \(\hat{\delta}_{\text{load}} = 73\) and \(\mu = 3.24\). The \AM{data-driven} model predictions closely resemble those of the analytical model, even for \( n = 0.2 \), despite the absence of training data in this parameter range. However, a closer look at the results for \( n = 0.2 \) reveals deviations from the analytical predictions provided by XPB. This indicates that while the ML model generalizes to unseen data, its accuracy weakens in regions where no prior information is available. In fact, this observation motivated us to incorporate physics-based augmentation to enhance the model's reliability in extrapolated scenarios.

Table~\ref{tab:PA_S1_results} summarizes the performance of 
ML model on the physics augmented data for four algorithms, where XPB outputs are used as additional input features. The MSE and $\textrm{R}^2$ are evaluated as summarized in Table ~\ref{tab:PA_S1_results}. The results demonstrate that tree-based models significantly outperform linear regression in terms of both accuracy (lower MSE and higher $\textrm{R}^2$) and robustness (lower standard deviation). \AM{This time, Random Forest} achieves the best performance. \AM{XGBoost} and Regression Tree models also show comparable performance, with only marginal differences in MSE and $\textrm{R}^2$ values, though the Random Forest exhibits substantially larger model sizes. The results in Table~\ref{tab:PA_S1_results} are further supported by Figure \ref{fig:PAML_pre_vs_gt} in \ref{app:ML}.

Comparing Table \ref{tab:PA_S1_results}, which presents the performance of the PA-ML approach, with the pure data-driven ML results in Table \ref{tab:ML_results} reveals a \AM{significant} reduction \AM{of 60.3\%} in the mean of MSE across the folds. However, note that the primary goal of physics augmentation is not merely to reduce error on seen data but to enhance generalization beyond unseen data, as illustrated in Fig. \ref{fig:ML_surface}(a) and (b). The best ML model in Table \ref{tab:ML_results} is XGBoost with a size of \AM{4993}, while the corresponding PA-ML approach in Table \ref{tab:PA_S1_results} has a size of \AM{4636}, resulting in a \AM{7.15}\% reduction. This trend is also observed in Regression Tree model, which shows size reductions of \AM{0.22\%. However, for the PA-ML models the best performance is achieved by Random Forest, while this performance is achieved at the cost of the increase in the model size.}

The key aspect of Figure \ref{fig:ML_surface} is that the XPB model remains valid in this region, enabling a meaningful comparison between the purely data-driven ML approach and the PA-ML approach. This comparison is illustrated in Figures \ref{fig:ML_surface}(a) and (b), where the circular data points represent ML predictions, and the dashed points correspond to XPB results. In regions with sufficient training data (n = 0.6 and n = 1.6, particularly at high unloading rates), both models align well with XPB. However, deviations become more pronounced in areas with data scarcity. Notably, the red data points correspond to n = 0.2, a region where no training data is available \AM{; yet, the ML model gives a fairly good representation of system behavior}. In this challenging regime, the purely data-driven model (Figures \ref{fig:ML_surface}(a)) exhibits deviations from XPB, while the PA-ML model (Figures \ref{fig:ML_surface}(b)) produces smoother and more accurate predictions, demonstrating its superior generalization beyond the training set. The PA-ML model predictions closely align with the XPB analytical solution at \( k = 0.1 \), the most representative value in the training set (see \ref{app:descriptive_stats}). We further evaluated the model against XPB across the range \( k = 0.5 \) to \( k = 0.02 \); nevertheless, it reliably captures the overall trend down to \( k = 0.02 \).

\AM{To gain insights into how variations in physical parameters influence the computational cost of numerical simulations, and highlight the computational efficiency offered by ML models, we present representative cases from the dataset. Using our BEM-based numerical simulations implemented in MATLAB 2023b on a desktop computer equipped with Windows 11 pro, a 12th Gen Intel(R) Core(TM) i9-12900K, 3200 Mhz, 16 Cores, and 96 GB RAM, the CPU computation time exhibits significant variability depending on the complexity of the physical inputs. For instance, evaluating cases such as $\mu=0.3$, $n=1.6$, $k=0.1$, $\widehat{r}_u=10^{10}$, and $\widehat{\delta}_l=12.87$ required approximately 113 seconds, whereas the computational effort drastically increased to over 35,385 seconds (approximately 9.8 hours) for a scenario with $\mu=3.24$, $n=0.22$, $k=0.1$, $\widehat{r}_u=10^{2}$, and $\widehat{\delta}_l=6.96$. In contrast, predictions generated by our trained PA-ML models, evaluated in Python 3.12 on a laptop with an Intel i7-6700HQ CPU, 16\,GB RAM, running Windows~10, remained consistently below 5 milliseconds per inference across all explored parameter combinations. This demonstrates that ML models provide nearly instantaneous predictions. Hence, ML-based models offer computational efficiency independent of the complexity inherent in the input parameters.
}

\begin{table}[h!]
\caption{Performance of PA-ML approaches for the prediction of effective energy surface: MSE values (scaled by $10^{-3}$) are presented asmean ± standard deviation, along with $\textrm{R}^2$ values obtained from 5-fold cross-validation and model sizes.}   \label{tab:PA_S1_results}
    \centering
    \small
    \begin{tabular}{lccc}
        \toprule
         & \textbf{MSE ($\cdot 10^{-3}$)} & \textbf{$\textrm{R}^2$} & \textbf{model size} \\
        \midrule
Linear Regression & $16.7658 \pm 1.0776$ & $0.8586 \pm 0.0102$ & 7\\
Regression Tree & $0.1355 \pm 0.0796$ & $0.9989 \pm 0.0007$ & 6443\\
XGBoost & $0.0977 \pm 0.0824$ & $0.9992 \pm 0.0007$ & 4636\\
Random Forest & $0.0575 \pm 0.0247$ & $0.9995 \pm 0.0002$ & 417800\\

        \bottomrule
    \end{tabular}
 
\end{table}

\begin{figure}[h] 
     \centering
      \includegraphics[width=0.49\textwidth]{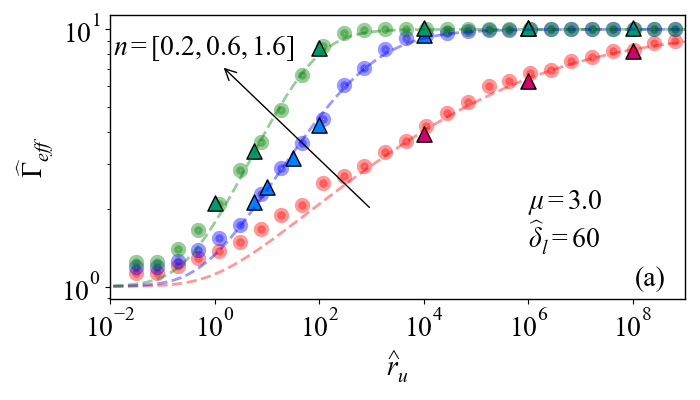}
     \includegraphics[width=0.49\textwidth]{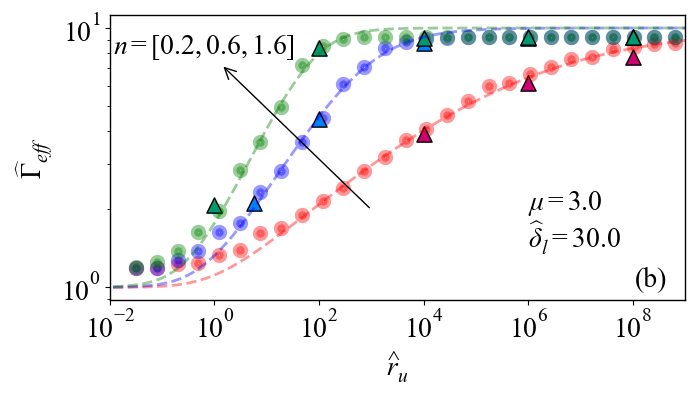}
     \includegraphics[width=0.49\textwidth]{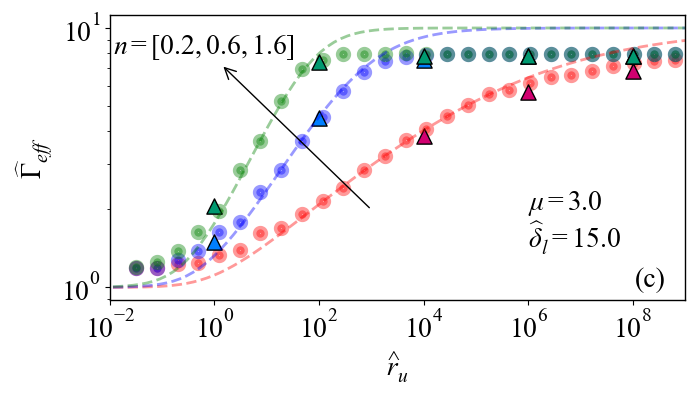}
     \includegraphics[width=0.49\textwidth]{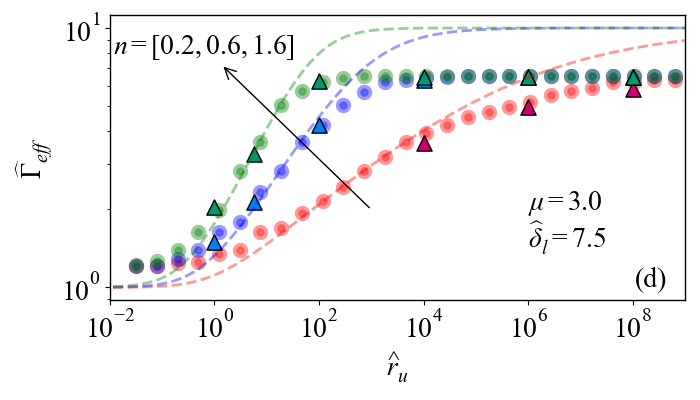} 
\caption{
Impact of indentation depth on rate-dependent effective surface energy, evaluated at a fixed Tabor parameter ($\mu = 3.0$), material modulus ratio ($k = 0.1$), and for three values of the power-law exponent $n = [0.2,\, 0.6,\, 1.6]$. 
Predictions from the PA-ML model are shown as circular markers, XPB analytical results as dashed lines, and test BEM results are represented by triangular markers. 
Subplots correspond to: (a) $\widehat{\delta}_l = 60$; (b) $\widehat{\delta}_l = 30$; (c) $\widehat{\delta}_l = 15$; and (d) $\widehat{\delta}_l = 7.5$.
}
 \label{fig:PA-ML_surface_S1}
\end{figure}

What we have accomplished so far is the development of a model that performs relatively well in regions where numerical results are unavailable. However, for the region plotted in Figure \ref{fig:ML_surface}, we have the XPB model with its very good performance. To understand the advantages of the ML models over the XPB model, one should examine its results in regions \AM{where XPB does not apply,} particularly where \(\mu\) and \(\widehat{\delta}_l\) are low. According to Figure \ref{fig:PA-ML_surface_S1}, XPB is only valid for high values of the Tabor parameter and high indentation depth (as seen in Figure \ref{fig:PA-ML_surface_S1}(a)). In contrast, Figure \ref{fig:PA-ML_surface_S1}(b), (c) and (d) clearly indicate that XPB fails at higher unloading rates, whereas the PA-ML results remain well-aligned with test data (triangular nodes obtained through BEM). By comparing Figures \ref{fig:PA-ML_surface_S1}(a) to (d), where the indentation depth decreases from $\widehat{\delta}_l = 60$ to $\widehat{\delta}_l = 7.5$, it is evident that the maximum viscoelastic amplification decreases as the indentation depth is reduced. Consequently, ML models serve as a valuable intermediary, particularly in regimes where analytical models (XPB) prove inadequate and numerical models (BEM) become computationally prohibitive. While XPB \AM{becomes not accurate outside} its valid range, BEM, despite its accuracy, becomes excessively expensive and impractical in certain parameter regimes due to exponential increases in computational cost and numerical instabilities at extreme unloading rates or indentation depths. 

\begin{figure}[h!] 
     \centering
      \includegraphics[width=0.49\textwidth]{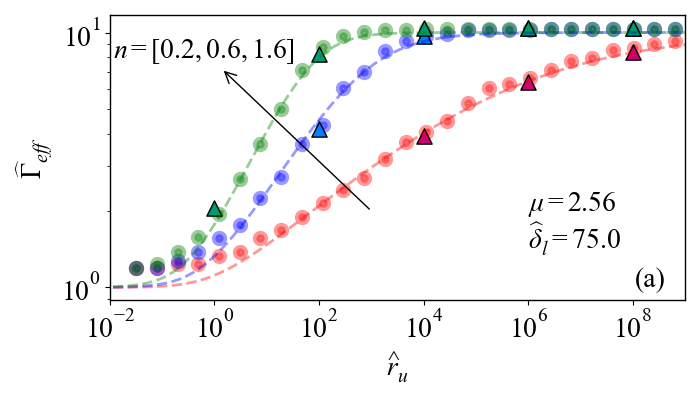}
     \includegraphics[width=0.49\textwidth]{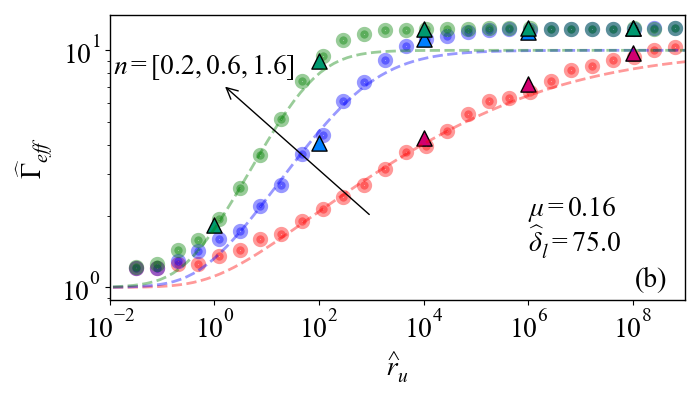}
     \includegraphics[width=0.49\textwidth]{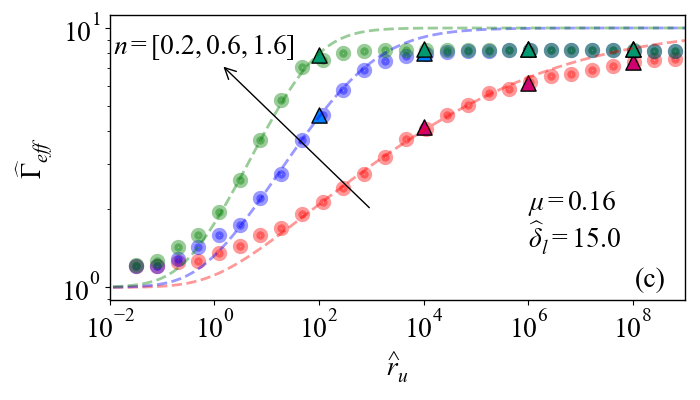}
     \includegraphics[width=0.49\textwidth]{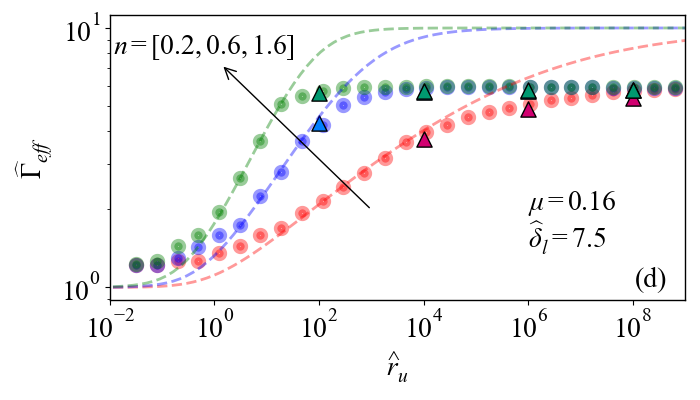} 
              \caption{The transition from JKR to DMT behavior captured by the PA-ML model for varying Tabor parameter values material modulus ratio ($k = 0.1$), and for three values of the power-law exponent $n = [0.2,\, 0.6,\, 1.6]$. Circular dots represent the PA-ML predictions, the dashed lines are the XPB solution, and the triangular nodes denote the test BEM results. \AM{Subplots correspond to: (a) $\mu = 2.56$, $\widehat{\delta}_l = 75$; (b) $\mu=0.16$  $\widehat{\delta}_l = 75$; (c) $\mu=0.16$ and $\widehat{\delta}_l = 15$; and (d) $\mu=0.16$, $\widehat{\delta}_l = 7.5$.}
              } \label{fig:PA-ML_surface_S2}
\end{figure}

We investigate the effect of varying the Tabor parameter \AM{on effective surface energy as a function of $\widehat{r}_u$ }for a constant indentation depth to evaluate the performance of PA-ML and compare its results with those of XPB, as shown in Figure \ref{fig:PA-ML_surface_S2}. The XPB predictions is represented by dashed lines, and the BEM results are shown by triangular nodes as test data. For an indentation depth of $\widehat{\delta}_l = 75$ and Tabor parameter values ranging from $\mu = 2.56$ to \AM{$\mu = 0.16$} in Figures \ref{fig:PA-ML_surface_S2}(a) to (b), a transition from JKR to DMT behavior is observed. This transition is not captured by the XPB model but is accurately detected by the PA-ML model. \AM{Also, Figures \ref{fig:PA-ML_surface_S2}(c) and (d) show that the model could consider the indentation depth effect in the DMT regime.}

\begin{figure}[H] 
     \centering
      \includegraphics[width=0.92\textwidth]{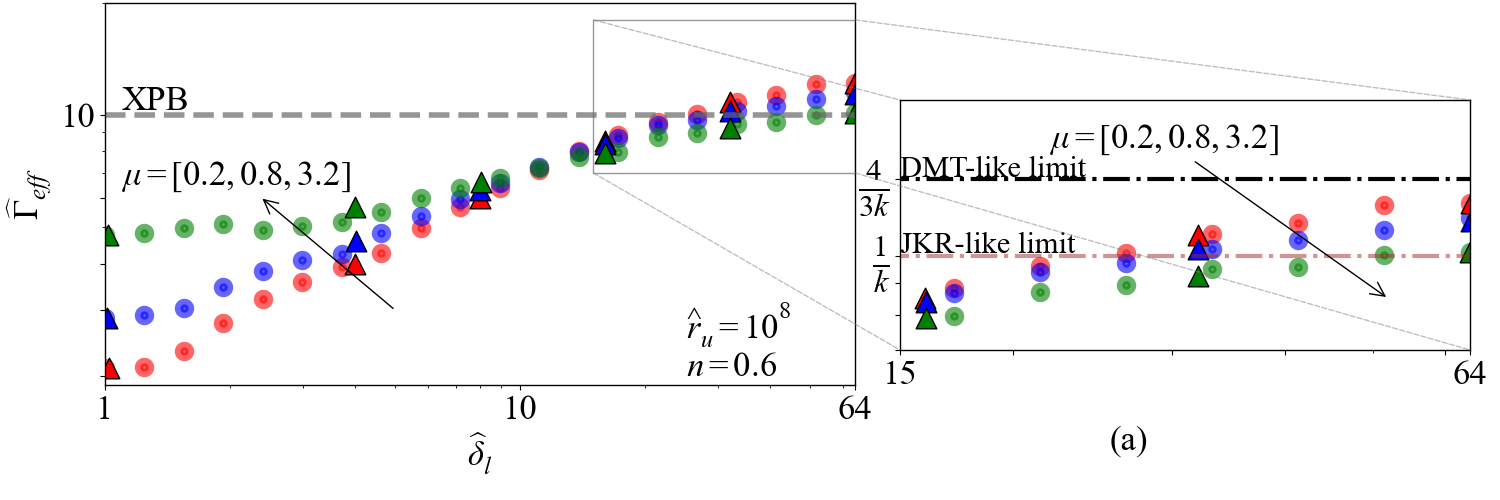}
      
     \includegraphics[width=0.55\textwidth]{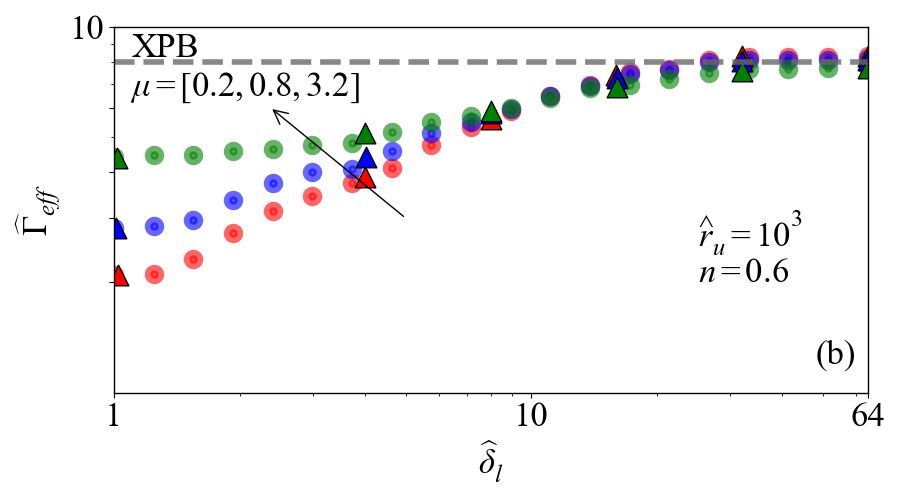}
     
    \includegraphics[width=0.55\textwidth]{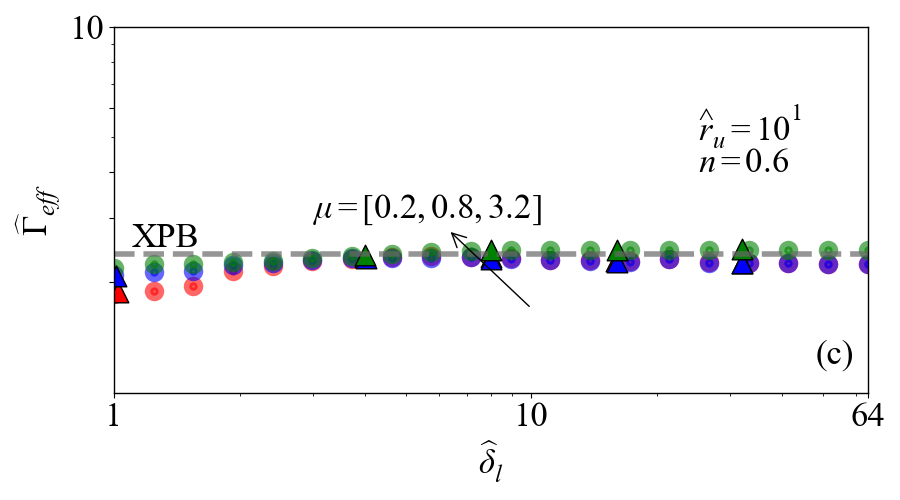}
      
              \caption{PA-ML model (Random Forest): Normalized surface energy versus indentation depth for different values of $\mu=[0.05,\,0.2,\,0.8,\,3.2]$. Dash-dotted lines represent the JKR limit (which is equal to $\frac{1}{k}$)\cite{maghami2024bulk} and DMT limit (which is $\frac{4}{3k}$) \cite{ciavarella2022upper}, gray dashed line indicates the XPB results, circular dots correspond to the BEM solution, and the solid lines indicate the PA-ML predictions for (a) $\widehat{r}_u=10^8$, $n = 0.6$, $k=0.1$; (b) $\widehat{r}_u=10^3$, $n = 0.6$, $k=0.1$; (c) $\widehat{r}_u=10^1$, $n = 0.6$, $k=0.1$ . The results indicate that the JKR and DMT limits were only achieved with high indentation depth values and very high unloading rates.} \label{fig:PA-MLdeltaeffect}
\end{figure}

Furthermore, we leveraged our PA-ML model to provide more insight into the interdependence of indentation depth ($\widehat{\delta}_l$) and the Tabor parameter ($\mu$), as illustrated in Figure~\ref{fig:PA-MLdeltaeffect}. In Figure~\ref{fig:PA-MLdeltaeffect}, the black dashed-dotted line represents the upper bound for DMT-like behavior, which is equal to $\frac{4}{3k} = 40/3$ \cite{ciavarella2022upper, wang2025rapid}, while the JKR-like limit ($1/k = 10$) is indicated by the mild brown dashed-dotted line. The XPB prediction is depicted by the dashed gray line, demonstrating its independence from variations in $\mu$ and $\widehat{\delta}_l$. We present the results for a very high unloading rate ($\widehat{r}_u = 10^9$) and a specific value of the material power law exponent ($n = 0.6$) in Figure~\ref{fig:PA-MLdeltaeffect}(a), which illustrates the highest achievable adhesion as a function of indentation depth and $\mu$. This point was briefly addressed in our previous study \cite{maghami2024bulk}, where it was noted that achieving adhesion is not solely a function of the unloading rate; the indentation depth also plays a significant role in determining the effective surface energy and, consequently, the adhesion. Here, we demonstrate the interdependence of $\widehat{\delta}_l$ and $\mu$ through our PA-ML predictions. It is important to note that for all cases similar to Figure~\ref{fig:PA-MLdeltaeffect}(a), where $k = 0.1$, $n = 0.6$, and the unloading rate is $\widehat{r}_u = 10^9$, the XPB model consistently predicts a value equal to the JKR limit of $1/k = 10$. 

From Figure~\ref{fig:PA-MLdeltaeffect}(a), it can also be deduced that for lower values of indentation depth, an increase in $\mu$ results in a growth in the effective surface energy. Conversely, for higher values of indentation depth, an increase in the $\mu$ parameter leads to a reduction in the effective surface energy, transitioning from a DMT-like limit to a JKR-like limit. Hence, one can deduce that the Tabor’s effect is not uniform, which means the depth-dependent Tabor effect on adhesion.

Considering Figure~\ref{fig:PA-MLdeltaeffect}(b), where we plotted the results for mid values of the unloading rate, it can be deduced that all scenarios corresponding to different values of $\mu$ for high values of indentation depth converge to a plateau equal to the XPB results, which is lower than both the JKR-like and DMT-like limits. Additionally, it is evident that for lower values of indentation depth, the dependency on the Tabor parameter persists, and for higher values of $\mu$, our PA-ML models predict higher adhesion. Furthermore, Figure~\ref{fig:PA-MLdeltaeffect}(c) shows that at low unloading rates (here, $\widehat{r}_u=10$), the effective surface energy remains almost independent of indentation depth and the Tabor parameter, yielding results aligned with the XPB model predictions.
\subsection{Results for prediction of work to pull off}
\label{sec:work}
\begin{figure}[h] 
     \centering
      \includegraphics[width=0.49\textwidth]{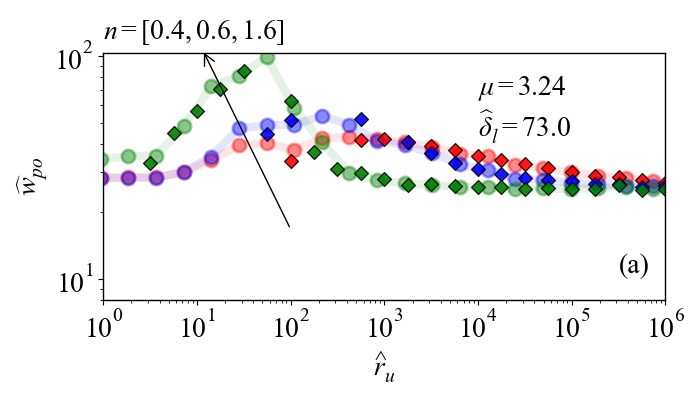}
     \includegraphics[width=0.49\textwidth]{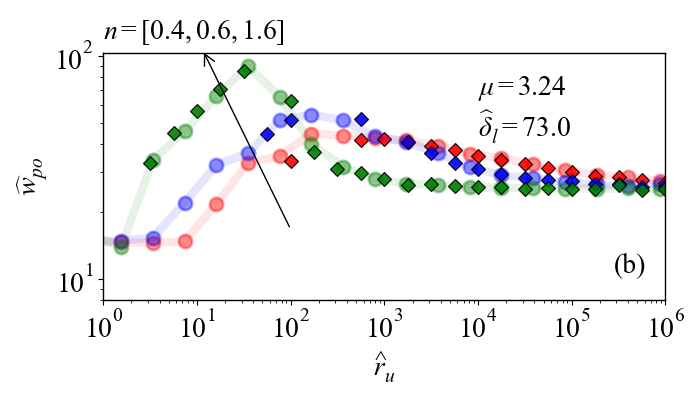}
      
\caption{
Comparison of (a) a purely data-driven ML model (Random Forest) and (b) a physics-augmented ML model (PA-ML Random Forest) in predicting the relationship between normalized work to pull-off and unloading rate. 
Diamonds represent the training data, circles denote model predictions. 
Results are evaluated for $\mu = 3.24$, $\widehat{\delta}_{\text{load}} = 73$, $k = 0.1$, and power-law exponents $n = [0.4,\, 0.6,\, 1.6]$.} \label{fig:work_ML}
\end{figure}

To gain insight into the energy required to detach a sphere from a viscoelastic surface, one can refer to the concept of work to pull off  \AM{$
w_{\text{po}} =  \int_{{\delta}_{\text{on}}}^{{\delta}_{\text{po}}} {P}(\delta, t) \, d{\delta} ={\widehat{w}_{po}}{(1.5 \pi \Delta\gamma_0Rh_0)},$
where ${\delta}_{\text{on}}$ is the displacement at which the normal force first becomes zero during unloading, and ${\delta}_{\text{po}}$ denotes the displacement at pull-off \cite{violano2022size} as shown in Figure \ref{fig:load_vs_disp}.} We employed the same approach and architecture of surrogate ML models used for predicting effective surface energy, but adapted it for the prediction of normalized work to pull off. For the pure data-driven ML approach, we again utilized five inputs, as outlined in Figure \ref{fig:fig_tab_data}, but with $\widehat{w}_{po}$ as the output representing the work to pull off. The performance results of the ML models are presented in \AM{Tables \ref{tab:ML_results_W} and \ref{tab:PAML_results_W}}. Consistent with the findings in Tables~\ref{tab:PA_S1_results}, Random Forest exhibits the best performance among the algorithms for predicting the work to pull off, achieving $\textrm{R}^2$ of $0.9945 \pm 0.0025$ across five folds. Consequently, we illustrate the prediction results of the work to pull off through Random Forest for a wide range of unloading rates and power law exponents in Figure~\ref{fig:work_ML}. 

\begin{table}[h!]
    \caption{Performance of ML approaches for the prediction of work to pull off ($\widehat{w}_{po}$): MSE and $\textrm{R}^2$ values (mean ± std. deviation) obtained from 5-fold cross-validation, with MSE reported as $\cdot 10^{-3}$ and model sizes included.} \label{tab:ML_results_W}
    \centering
    \small
    \begin{tabular}{lccc}
        \toprule
         & \textbf{MSE ($\cdot 10^{-3}$)} & \textbf{$\textrm{R}^2$} & \textbf{model size} \\
        \midrule
Linear Regression & $51.4637 \pm 12.5943$ & $0.9062 \pm 0.0227$ & 6\\
Regression Tree & $4.6163 \pm 2.0924$ & $0.9916 \pm 0.0038$ & 6503\\
XGBoost & $3.1094 \pm 1.8688$ & $0.9943 \pm 0.0035$ & 4766\\
Random Forest & $2.9948 \pm 1.3544$ & $0.9945 \pm 0.0025$ & 419700\\
        \bottomrule
    \end{tabular}
\end{table}

\begin{table}[h!]
    \caption{Performance of PA-ML approaches for the prediction of work to pull off ($\widehat{w}_{po}$): MSE and $\textrm{R}^2$ values (mean ± std. deviation) obtained from 5-fold cross-validation, with MSE reported as $\cdot 10^{-3}$ and model sizes included.} \label{tab:PAML_results_W}
    \centering
    \small
    \begin{tabular}{lccccc}
        \toprule
         & \textbf{MSE ($\cdot 10^{-3}$)} & \textbf{$\textrm{R}^2$} & \textbf{model size} \\
        \midrule
Linear Regression & $51.1512 \pm 12.5556$ & $0.9067 \pm 0.0226$ & 7\\
Regression Tree & $4.0330 \pm 2.0619$ & $0.9926 \pm 0.0037$ & 6493\\
XGBoost & $2.9248 \pm 1.8478$ & $0.9946 \pm 0.0034$  & 4954\\
Random Forest & $2.4197 \pm 1.2864$ & $0.9956 \pm 0.0023$  & 419400\\

        \bottomrule
    \end{tabular}
\end{table}

Figure~\ref{fig:work_ML} demonstrates that the work to pull off \AM{versus the unloading rate} exhibits a different behavior compared to the effective surface energy. Unlike the effective surface energy, the work to pull off follows a bell-shaped curve as a function of the unloading rate \AM{as also have been shown in a work by full numerical modeling \cite{violano2022size}.} The results in Figure~\ref{fig:work_ML} reveal an interacting effect between the unloading rate and the power law exponent on the work to pull off, which was not observed in the results for the effective surface energy. Specifically, for relatively lower values of the unloading rate, an increase in the power law exponent leads to a rise in the work to pull off, whereas for higher values of the unloading rate, our PA-ML model predictions indicate that an increase in the power law exponent results in a decrease in the work to pull off. 

It is important to note that, to date, there has been no analytical model capable of describing the work to pull off the concept across any range of material variables. Therefore, for physical augmentation in our PA-ML and the prediction of work to pull off, we decided to utilize the XPB model's output. In the following results, we aim to evaluate the effect of incorporating the outputs of the XPB model—specifically, the effective surface energy—on the prediction of a distinct parameter, the work to pull off.

 According to our results, \AM{Random Forest performs better than the} other algorithms in terms of MSE, $\textrm{R}^2$. The most notable observation from Table~\ref{tab:PAML_results_W} is the significant impact of augmentation through the effective surface energy ($\widehat{\Gamma}_{eff}$ based on XPB) on the performance of algorithms in predicting the work to pull off ($\widehat{w}_{po}$). A comparison between Table~\ref{tab:PAML_results_W} and Table~\ref{tab:ML_results_W} reveals the influence of incorporating this physical parameter as an additional input into the modeling process. 
 The results indicates that augmentation consistently reduces prediction errors and enhances performance, as evidenced by improvements in both MSE and $\textrm{R}^2$. Additionally, the model sizes of \AM{Random Forest} reported in Table \ref{tab:ML_results_W} and \ref{tab:PAML_results_W} suggest that the integration of XPB not only improves predictive accuracy but also leads to more efficient and compact models.

\begin{figure}[h] 
     \centering
      \includegraphics[width=0.49\textwidth]{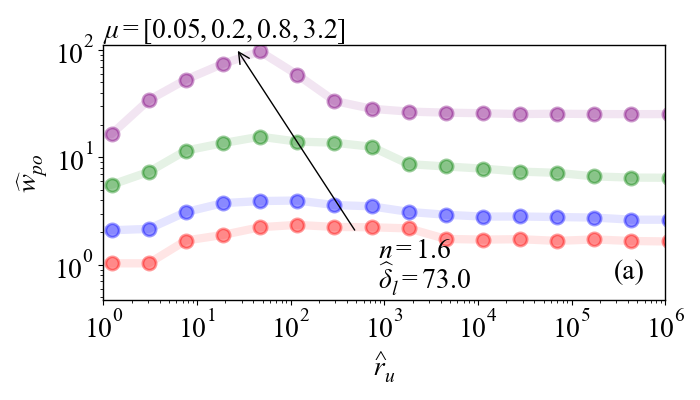}
     \includegraphics[width=0.49\textwidth]{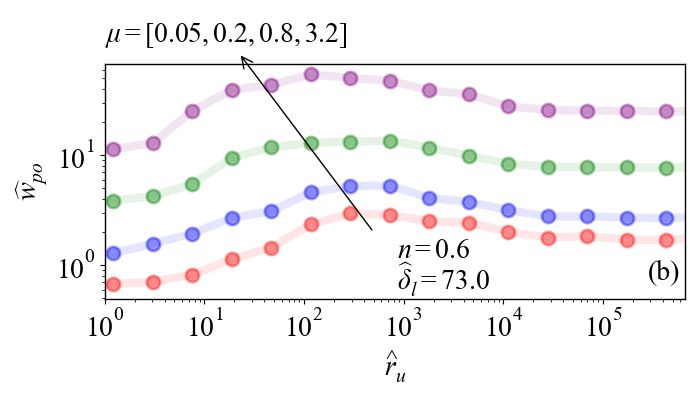}
      
             \caption{
PA-ML model (Random Forest): Normalized work to pull-off as a function of unloading rate for different values of the Tabor parameter ($\mu = [0.05,\, 0.2,\, 0.8,\, 3.2]$), at a fixed indentation depth $\widehat{\delta}_l = 73.0$ and material modulus ratio $k = 0.1$. 
Circular markers represent predictions from the PA-ML model. Results are shown for two power-law exponents: (a) $n = 1.6$ and (b) $n = 0.6$.}
 \label{fig:work_PAML}
\end{figure}

To visually assess the effect of XPB for guiding the PA-ML model \AM{for prediction of work to pull off}, we illustrate the predictions of \AM{Random Forest} across a wide range of unloading rates and varying values of the power law exponent in Figure~\ref{fig:work_ML}(b). This figure is structured similarly to Figure~\ref{fig:work_ML}(a), where the circles represent the PA-ML predictions, and the diamonds denote the training data. A comparison between Figures~\ref{fig:work_ML}(a) and~\ref{fig:work_ML}(b) reveals that the PA-ML model provides predictions that are closer to the BEM solutions and demonstrate smoother behavior.

To gain insight into different slices of the material behavior space and to examine the effects of input parameters on the material's behavior, we leveraged our PA-ML model to explore the effect of the Tabor parameter ($\mu$) in Figure~\ref{fig:work_PAML}. The results are presented as a function of unloading rate and two specific values of the material power law exponent ($n = 1.6$) in Figure~\ref{fig:work_PAML}(a), as well as for $n=0.6$ in Figure~\ref{fig:work_PAML}(b). The work to pull off consistently increases as $\mu$ increases in both Figure~\ref{fig:work_PAML}(a) and Figure~\ref{fig:work_PAML}(b). To ensure the validity of the ML predictions, we provide analytical estimations of the upper and lower bounds for the work to pull off in \ref{app:up} and \ref{app:low}, respectively.

\section{Summary and conclusions} \label{sec:conc}
We introduced \AM{two} ML-based approaches for predicting the \AM{pull off force and work to pull off} of relaxed viscoelastic \AM{adhesive} Hertzian contacts \AM{as a function of five input parameters of Tabor parameter $\mu$, material power exponent $n$, modulus ratio $k$, unloading rate $\widehat{r}_u$, and the inidentation depth $\widehat{\delta}_l$}. Based on \AM{8921} samples generated through our BEM numerical framework, the pure data-driven ML models efficiently provide predictions in regions where computational costs are high. Notably, their strong agreement with our previously developed analytical model further validates their reliability and effectiveness. This confirms that the proposed approach could be seen as a bridge between analytical models and numerical methods. Through a systematic comparison, the study demonstrates that ML-based approaches, particularly tree-based methods like \AM{Random Forest}, excel in predicting tabular data with low MSE and high \(\textrm{R}^2\) values. The integration of ML with physics-based insights, particularly guided by the XPB model, developed by Maghami et al \cite{maghami2024bulk}, \AM{enables} the creation of efficient and accurate surrogate models for predicting key adhesion-related quantities, including effective surface energy and work to pull off, \AM{especially where analytical models fail and numerical simulations become costly.} Notably, the PA-ML models not only improve prediction accuracy but also, \AM{help generalization.} Despite the distinct relationship between the work to pull off quantity and effective surface energy, integrating XPB guidance, which provides the latter in a \AM{valid} region, enhanced the prediction accuracy of the former overall. The PA-ML framework helped us to reveal interesting insights into the adhesion mechanics of viscoelastic materials. Particularly, the interplay between the Tabor parameter and indentation depth was clarified, showing a transition between DMT-like and JKR-like behaviors under different conditions. A depth-dependent Tabor effect on adhesion is detected by our PA-ML. For lower indentation depths, increasing the Tabor parameter led to a rise in effective surface energy. Conversely, at higher indentation depths, the behavior transitioned to a JKR-like regime, resulting in lower adhesion. PA-ML prediction showed that the work-to-pull-off exhibited different behavior from the effective surface energy, following a bell-shaped curve as a function of the unloading rate. The interacting effects of the unloading rate and the power-law exponent were observed, revealing that the work-to-pull-off increases with the power-law exponent at lower unloading rates but decreases at higher unloading rates. 
Overall, the results demonstrate that the PA-ML framework serves as a valuable intermediary between analytical and numerical methods, addressing limitations in generalization, computational cost, and numerical instability. While analytical models like XPB provide theoretical consistency and numerical methods like BEM offer precision, the PA-ML approach combines the strengths of both to offer interpretable and computationally efficient predictions. The presented framework lays the foundation for further advancements \AM{in the data driven approachesd to model visco-adhesive contact problems.}
\pagebreak

\section*{Acknowledgments}
This work was accomplished during a research stay of the first author (Ali Maghami) at the Chair of Cyber-Physical Systems in Mechanical Engineering of the Technische Universität Berlin. A.M., and A.P. were partly supported by the Italian Ministry of University and Research under the Programme
‘‘Department of Excellence’’ Legge 232/2016 (Grant No. CUP - D93C23000100001). A.P., and A.M. were supported by the European Union (ERC-2021-STG, ‘‘Towards Future Interfaces With Tuneable Adhesion By Dynamic Excitation’’ - SURFACE,
Project ID: 101039198, CUP: D95F22000430006). Views and opinions expressed are however those of the authors only and do not
necessarily reflect those of the European Union or the European Research Council. Neither the European Union nor the granting
authority can be held responsible for them. A.P. was partly supported by the European Union through the program – Next Generation
EU (PRIN-2022-PNRR, ‘‘Fighting blindness with two photon polymerization of wet adhesive, biomimetic scaffolds for neurosensory
REtina-retinal Pigment epitheliAl Interface Regeneration’’ - REPAIR, Project ID: P2022TTZZF, CUP: D53D23018570001). 

\section{Data availability}
\AM{The dataset, and the ML models generated for this article is available on Zenodo at \\ doi.org/10.5281/zenodo.15039240, and a GitHub repository at github.com/alimaghamii/ML4Adhesion.
}
\appendix
\section{Details of the boundary element method implementation as a numerical model}\label{app:BEM}
\subsection{{Kernal function}}
$G\left(  r,s\right)$ is the so-called Kernel function, defined as:%

\begin{equation}
G\left(  r,s\right)  =\left\{
\begin{array}
[c]{cc}%
\frac{4}{\pi r}K\left(  \frac{s}{r}\right)  , & \qquad\qquad s<r\\
\frac{4}{\pi s}K\left(  \frac{r}{s}\right)  , & \qquad\qquad s>r
\end{array}
\right.  \label{kernel}%
\end{equation}
where \(K(k)\) denotes the complete elliptic integral of the first kind with modulus \(k\). 

In this work, the gap function is solved applying the Boundary Element Method (BEM) combined with the Newton–Raphson method on \( N = M + 1 \) equally spaced nodes, where \( M \) represents the number of interfacial elements.  Equation (\ref{eq:uz}) is discretized in both time and space to compute the half-space deflections. A time-marching algorithm with a time step \( \Delta t \) is employed for the temporal discretization. For spatial discretization, the pressure distribution is assumed to have a triangular shape over each element. Specifically, for the \( j\)-th element, the pressure \( p_j \) is defined at \( r = r_j \) and decreases linearly to zero at \( r = r_{j-1} \) and \( r = r_{j+1} \). This approach is commonly referred to as the \emph{method of overlapping triangles} \cite{johnson1987contact,papangelo2020numerical, papangelo2023detachment}. Once the number of nodes is fixed the influence matrix $G_{i,j}$ can be computed so that the deflection $u_z[i]$ of the \textit{elastic} halfspace at the node $i$ can be found by linear superposition 

\begin{equation}
u_z[i] = \frac{1}{E^*_0}\sum_{j=1}^N G_{i,j}\sigma_{j}.
\end{equation}

For the problem of a Hertzian rigid indenter on a viscoelastic surface, one can define the following dimensionless parameters as: \( \widehat{h}=(h - h_0)/h_0 \), \( \widehat{\delta}=\delta / h_0  \), \( \widehat{r} = r/\beta  \), $
\widehat{\sigma} = \sigma /( \frac{\Delta \gamma_0}{\mu h_0} ) \; $, and ${t}=\tau \widehat{t}$ where 
\begin{equation}
\beta^3 = \frac{R^2 \Delta \gamma_0}{E^*}, \quad \mu = \left( \frac{R \Delta \gamma_0^2}{{E_0^*}^ 2 h_0^3} \right)^{1/3}, 
\end{equation}

Equations (\ref{eq:LJ}), (\ref{eq:h}), and (\ref{eq:uz}) are expressed in a discretized form in both space and time, where $i$ denotes the spatial index and $q$ represents the temporal index, as follows:

\begin{equation}
\widehat{\sigma}[i,q] = -\frac{8}{3}\mu \left[ \frac{1}{(\widehat{h}[i,q] + 1)^3} - \frac{1}{(\widehat{h}[i,q] + 1)^9} \right]\;,
\end{equation}

\begin{equation}
\widehat{h}[i,q] = -\widehat{\delta}[q] +  \frac{1}{2}\mu \widehat{r}[i]^2 + \widehat{u}_z[i,q]\;,
\end{equation}

\begin{equation}
\widehat{u}_z[i,q] \approx \mu\sum \widehat{G}_{ij}\sum_{m=0}^{q} \widehat{C}[q-m]
\left(\widehat{\sigma}[j,m+1] - \widehat{\sigma}[j,m])\right)\;,
\end{equation}
where the dimensionless form of the creep compliance function $\widehat{C}=C/C_0$ in (\ref{eq:Ctgen}) is:
\begin{equation}
\widehat{C}(\widehat{t}) = 1 - 2 \frac{(1 - k)}{\Gamma(n)} \widehat{t}^{n/2} \mathbf{K}_{n}\left(2 \sqrt{\widehat{t}}\right) \,.
\end{equation}\label{eq:noncreep}

\section{Machine Learning Methods} \label{app:ML}
\subsection{Linear regression}
Linear regression is one of the simplest and most widely used statistical methods for modeling the relationship between a dependent variable and one or more independent variables. It assumes a linear relationship between inputs and outputs. Linear regression is particularly effective when the relationship between variables is approximately linear, but it may struggle with non-linear relationships unless transformed appropriately \cite{molnar2025interpretable}.
\subsection{Regression tree}
A regression tree is similar to a decision tree that predicts continuous outputs by recursively partitioning the data space into regions with similar target values. Each split is chosen to minimize variance in the resulting subsets. Regression trees are interpretable but prone to overfitting and can be unstable with small data changes \cite{molnar2025interpretable}.
\subsection{Random forest}
Random Forest is a learning method that constructs multiple decision trees during training and outputs the mode or mean prediction of these trees for classification or regression tasks. This technique enhances predictive accuracy and controls overfitting by averaging out biases from individual trees. Each tree in the forest is built using a random subset of the data and features, which helps capture diverse patterns within the dataset. Random Forest is robust to noise and can handle large datasets with high dimensionality effectively, making it a popular choice for many practical applications \cite{molnar2025interpretable}.
\subsection{Extreme Gradient Boosting (XGBoost)}
XGBoost is an advanced implementation of gradient boosting that has gained popularity due to its high performance in ML competitions. It builds models sequentially by adding new trees that correct errors made by previous ones, optimizing for both speed and accuracy through techniques like regularization and parallel processing \cite{chen2016xgboost}. XGBoost is particularly effective for tabular data because it can handle missing values internally and offers flexibility in model tuning.
\subsection{Visual comparison of ground truth and predictions}
\begin{figure}[t] 
     \centering
      \includegraphics[width=0.49\textwidth]{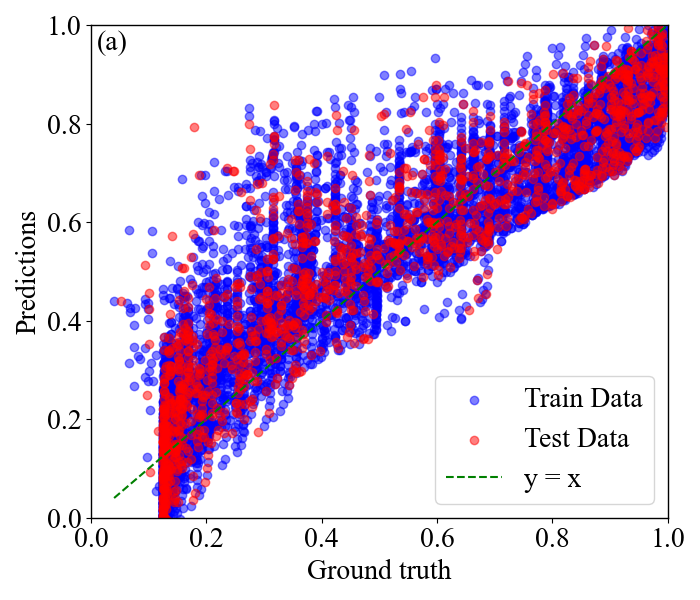}
      \includegraphics[width=0.49\textwidth]{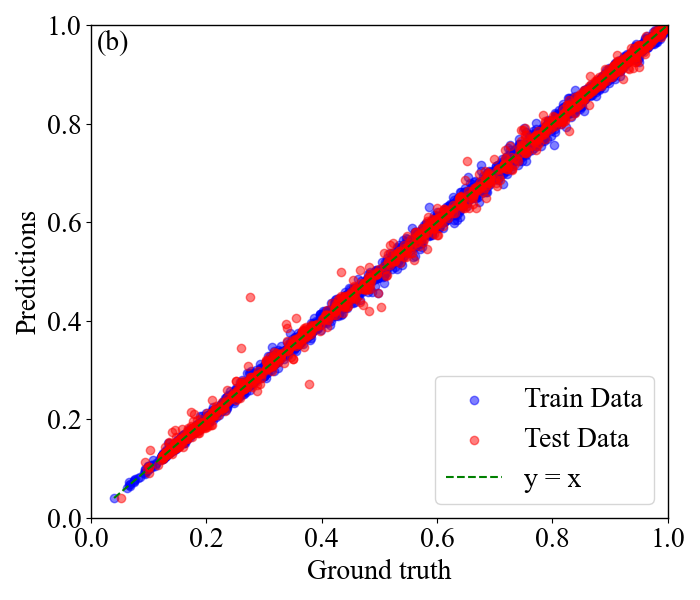}
              \caption{ML model's performance: (a) The plot illustrates the performance of linear regression model; (b) The plot shows the results of the XGBoost model, which demonstrates stronger performance.} \label{fig:ML_pre_vs_gt}
\end{figure}

\begin{figure}[t] 
     \centering
      \includegraphics[width=0.49\textwidth]{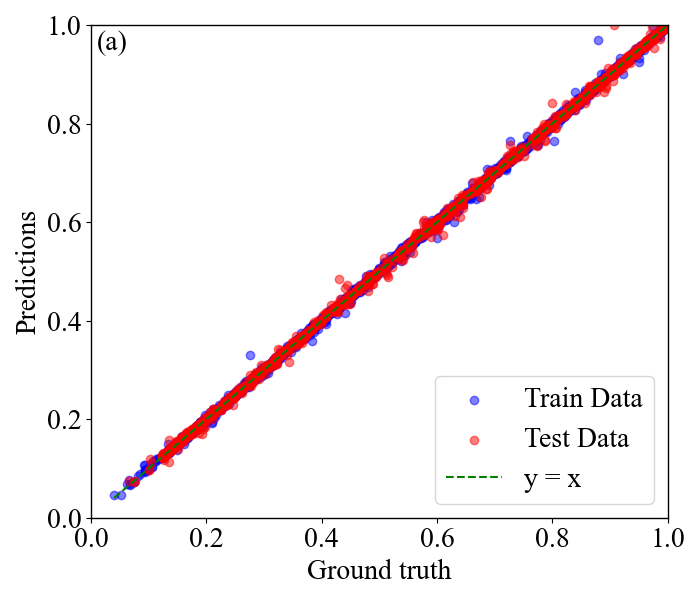}
      \includegraphics[width=0.49\textwidth]{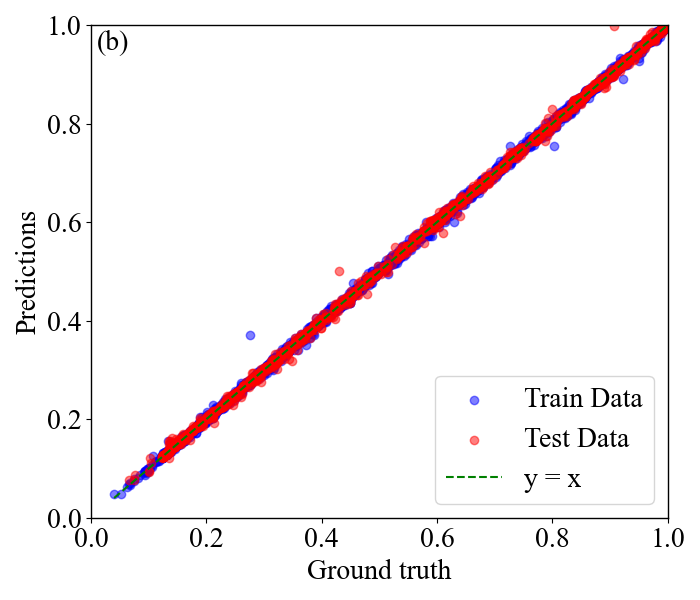}
              \caption{PA-ML model's performance: (a) The plot shows the performance of Random Forest; (b) The plot illustrates the results of the XGBoost model.} \label{fig:PAML_pre_vs_gt}
\end{figure}

Here, we present two figures that illustrate the performance of different modeling approaches. Figure \ref{fig:ML_pre_vs_gt} focuses on pure data-driven ML model predictions, while Figure \ref{fig:PAML_pre_vs_gt} shows the performance of a physics-augmented model. Each figure has two subfigures: the first subfigure depicts the results of linear regression, and the second represents the results from the XGBoost algorithm. 

\section{Descriptive Statistics of the Dataset}\label{app:descriptive_stats}

This section provides the descriptive statistics for the dataset. Table~\ref{tab:descriptive_stats} summarizes key statistical measures, including central tendency (mean, median), variability (standard deviation), and extreme values (minimum, maximum) for each parameter. 

The variables $\widehat{\delta}_l$, $\widehat{r}_u$, $k$, $\widehat{\Gamma}_{\text{eff}}$, and $\widehat{w}_{\text{po}}$ were $\log_{10}$-transformed before model training and testing. The minimum and maximum values indicate substantial heterogeneity within the dataset, particularly for $\widehat{r}_u$, which spans multiple orders of magnitude.

\begin{table}[h!]
 \caption{Descriptive statistics for the principal parameters in the data set.}
   \label{tab:descriptive_stats}
\centering
    \small
    \begin{tabular}{lccccc}
        \toprule
        \textbf{Statistic} 
        & $\log_{10}(\widehat{\delta}_l)$ 
        & $\mu$ 
        & $\log_{10}(\widehat{r}_u)$ 
        & $\log_{10}(k)$ 
        & n \\ 
        \midrule
        \textbf{Mean}     
        & 0.5718 
        & 0.7181 
        & 4.467  
        & -0.9189  
        & 0.8471  \\  
        \textbf{Median}   
        & 0.684  
        & 0.3   
        & 4.4    
        & -1     
        & 0.6  \\  
        \textbf{StdDev}   
        & 0.9359 
        & 0.9501 
        & 2.75   
        & 0.219  
        & 0.5635 \\
        \textbf{Min}      
        & -2.255 
        & 0.04   
        & -1.5   
        & -4     
        & 0.2 \\    
        \textbf{Max}      
        & 2.738  
        & 3.24   
        & 10     
        & -0.301 
        & 2 \\      
        \bottomrule
    \end{tabular}
\end{table}

\section{Upper-bound approximation of Hertzian pull-off work in the glassy Regime} \label{app:up}
At very high unloading rates, viscoelastic substrates behave as elastic solids characterized by their instantaneous (glassy) modulus. In this regime, adhesive detachment is dominated by elastic energy storage rather than viscoelastic dissipation. While the Hertzian geometry exhibits a continuously varying contact area, the stress distribution near detachment becomes increasingly concentrated toward the edge, resembling the uniform stress field under a flat punch. Therefore, we approximate the unloading behavior of an adhesive Hertzian contact at high rates using the flat punch formulation. In the limit of \textit{very fast unloading}, a viscoelastic substrate responds elastically with its \textit{glassy modulus} \( E^*_{\infty} \). For a rigid axisymmetric flat punch of radius \( a \) detaching from such a substrate, the normal force during unloading is as follows:
\begin{equation}
P(\delta) = 2aE^*_{\infty} \delta\,.
\end{equation}
The pull-off load is governed by linear elastic fracture mechanics and given by \cite{papangelo2023detachment}:
\begin{equation}
P_{\text{po}} = \sqrt{8\pi E^*_{\infty} \Delta\gamma_0 a^3}\,.
\end{equation}
Substituting this into the force-displacement relation yields the displacement at pull-off:
\begin{equation}
\delta_{\text{po}} = \frac{P_{\text{po}}}{2aE^*_{\infty}} = \frac{\sqrt{8\pi E^*_{\infty} \Delta\gamma_0 a^3}}{2aE^*_{\infty}}\,.
\end{equation}
The work required to pull off the punch is defined as the area under the unloading curve:
\begin{equation}
w_{\text{po}} = \int_0^{\delta_{\text{po}}} 2aE^*_{\infty} \delta \, d\delta = aE^*_{\infty} \delta_{\text{po}}^2\,,
\end{equation}
Substituting for \( \delta_{\text{po}} \) leads to the simplified expression:
\begin{equation}\label{eq:work_to_pull}
w_{\text{po}} = 2\pi a^2 \Delta\gamma_0\,,
\end{equation}
where it is the work to pull-off related to an axisymmetric flat punch, and for a Hertzian indenter, one can consider it with $a\approx a_{po}$. Considering the contact radius as $a=\widehat{a}{(\pi R^2 \Delta\gamma_0/E_0^*)^{1/3}}$, one can obtain the following relation:
\begin{equation}\label{eq:work_to_pull2}
w_{\text{po}} = 2\pi \widehat{a}_{\textrm{po}}^2{(\pi R^2 \Delta\gamma_0/E_0^*)^{2/3}} \Delta\gamma_0\,.
\end{equation}
Normalizing by the characteristic energy scale \(1.5\pi \Delta\gamma_0 R h_0\), the dimensionless work becomes:
\begin{equation}
\widehat{w}_{\text{po}} = \frac{w_{\text{po}}}{1.5\pi \Delta\gamma_0 R h_0}
= \frac{4}{3} \widehat{a}_{\textrm{po}}^2{\pi}^{2/3} \frac{R^{1/3} \Delta\gamma_0^{2/3}}{E^{2/3}_{0} h_0},
\end{equation}
Considering the Tabor parameter, we have the following upper bound approximation:
\begin{equation}\label{eq:work_to_pull3}
(\widehat{w}_{\text{po}})_{\textrm{up}}= \frac{4}{3} \widehat{a}_{\textrm{po}}^2 {\pi}^{2/3} \mu.
\end{equation}

\begin{figure}[h]
    \centering
 \includegraphics[width=0.49\textwidth]{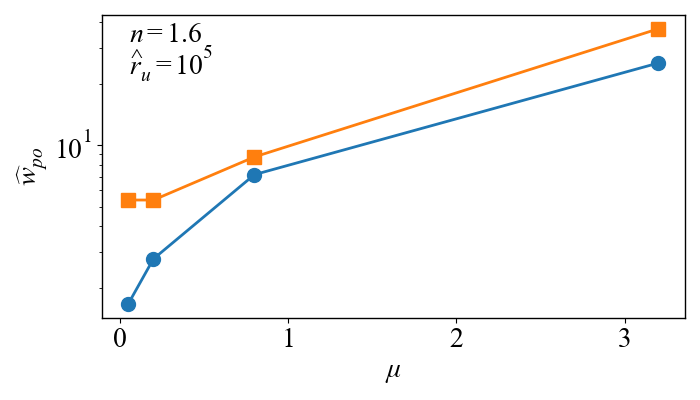} 
    \caption{
Comparison between the predicted normalized work to pull off ($\widehat{w}_{po}$) from the PA-ML model (XGBoost) and the upper bound of glassy behavior values obtained from equation~\eqref{eq:work_to_pull3}, for different values of the Tabor parameter ($\mu = [0.05,\, 0.2,\, 0.8,\, 3.2]$). 
The predictions correspond to a fixed normalized indentation depth $\widehat{\delta}_l = 73.0$, material modulus ratio $k = 0.1$, and normalized unloading rate $\widehat{r}_u = 10^5$ and $n = 1.6$. 
Circular markers indicate PA-ML predictions, and square markers correspond to values from equation~\ref{eq:work_to_pull3}.}
    \label{fig:work_vs_mu}
\end{figure}

Figure~\ref{fig:work_vs_mu} compares the PA-ML predictions and analytically estimated upper-bound of glassy behavior for normalized work to pull off from \ref{eq:work_to_pull3} across different Tabor parameters.

\section{Lower-bound approximation of work to pull off in JKR rubbery regime}\label{app:low}
\begin{figure}[h]
    \centering
    \includegraphics[width=0.6\textwidth]{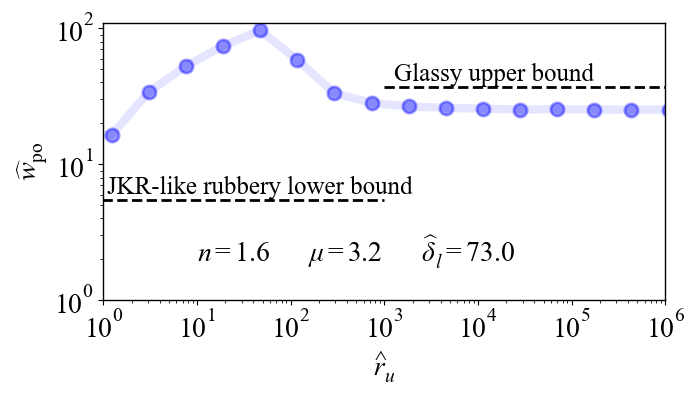}
    \caption{
    PA-ML prediction of normalized work to pull off ($\widehat{w}_{\mathrm{po}}$) as a function of the normalized unloading rate ($\widehat{r}_u$) for $\mu = 3.2$ and $n = 1.6$, at a fixed indentation depth $\widehat{\delta}_l = 73.0$ and modulus ratio $k = 0.1$. 
    The blue curve represents the PA-ML model prediction. 
    A horizontal dashed line marks the \textit{JKR-like rubbery lower bound} ($\widehat{w}_{\mathrm{po}} = 5.50376$ obtained through \eqref{eq:work_low}), which applies in the low-rate regime. 
    Another dashed line indicates the \textit{glassy upper bound} ($\widehat{w}_{\mathrm{po}} = 37.1203$ obtained through \eqref{eq:work_to_pull3}), relevant at higher unloading rates ($\widehat{r}_u \gtrsim 10^3$).
    }
    \label{fig:mu3.2_bounds}
\end{figure}

The Johnson-Kendall-Roberts (JKR) \cite{johnson1971surface} theory is an analytical model describing the mechanics of adhesive elastic contact. It extends Hertzian contact theory by considering surface energy, making it particularly applicable to soft materials and scenarios where adhesion plays a significant role. Here to have consistent relations with our numerical results as well as the ML solution, we consider $a=\widehat{a}  {\left(\pi R^{2} \Delta\gamma_0 / E_0^{*}\right)^{1/3}}$, and $P=\widehat{P} ({1.5{\pi} \Delta\gamma_0 {R}h_0})$. Hence, the JKR solution for the  dimensionless load and indentation are given by:
\begin{equation}
\widehat{P} = 2\widehat{a}^3 -1.5 \sqrt{8\widehat{a}^3}\;,
\end{equation}
\begin{equation}
\frac{\widehat{\delta}}{\mu\pi^{2/3}} = {\widehat{a}^2 - \sqrt{2\widehat{a}}}\;.
\end{equation}
The work to pull off is $
\widehat{w}_{\text{po}} = | \int_{\widehat{\delta}_{\text{on}}}^{\widehat{\delta}_{\text{po}}} \widehat{P}(\delta, t) \, d\widehat{\delta} |=\frac{w_{po}}{1.5 \pi \Delta\gamma_0Rh_0},$
where $\widehat{\delta}_{\text{on}}$ is the displacement at which the normal force first becomes zero during unloading, and $\widehat{\delta}_{\text{po}}$ denotes the displacement at pull-off. Therefore, the work to pull off is obtained as follows:
\begin{equation}\label{eq:work_low}
\widehat{w}_{\textrm{po}} = 0.80182 \mu\pi^{2/3}\;.
\end{equation}
As shown in Figure~\ref{fig:mu3.2_bounds}, the PA-ML model results in the glassy and rubbery regimes are bounded by the rubbery lower limit derived from Eq.~\eqref{eq:work_low} and the glassy upper limit given by Eq.~\eqref{eq:work_to_pull3}.

 \bibliographystyle{elsarticle-num}
 
 \bibliography{cas-refs}





\end{document}